%% file: bare_jrnl.tex
\documentclass[journal]{IEEEtran}
\ifCLASSINFOpdf
\else
\fi

\usepackage[ruled,vlined,linesnumbered]{algorithm2e}
\usepackage{amsmath,graphicx}
\usepackage{bm}
\usepackage{color}
\usepackage{amssymb}
\usepackage[hyphens]{url}
\usepackage{hyperref}
\hypersetup{
    colorlinks=true,
    linkcolor=blue,
    urlcolor=magenta,
}
\usepackage{mwe}
\usepackage{booktabs}
\usepackage[pagewise,switch]{lineno}
\usepackage[space]{cite}
\usepackage{float}

\usepackage{footnote}
\makesavenoteenv{tabular}

\newcommand{\lwlrap}{\emph{l$\omega$lrap}}
\newcommand{\rpm}{\raisebox{.2ex}{$\scriptstyle\pm$}}

\begin{document}
\title{FSD50K: An Open Dataset of \\ Human-Labeled Sound Events}

\author{Eduardo~Fonseca,~\IEEEmembership{Student Member,~IEEE,}
        Xavier~Favory,
        Jordi Pons,
        Frederic Font,
        and~Xavier~Serra% <-this % stops a space
\thanks{The authors are with the Music Technology Group at Universitat Pompeu Fabra, Barcelona, Spain. e-mail for all authors: (name.surname@upf.edu).}% <-this % stops a space
\vspace{-4mm}}


\markboth{IEEE/ACM Transactions on Audio, Speech, and Language Processing,~Vol.~30,~2022}%
{Shell \MakeLowercase{\textit{et al.}}: Bare Demo of IEEEtran.cls for IEEE Journals}



\maketitle

% The abstract must be between 150-250 words.
\begin{abstract}
Most existing datasets for sound event recognition (SER) are relatively small and/or domain-specific, with the exception of AudioSet, based on over 2M tracks from YouTube videos and encompassing over 500 sound classes.
However, AudioSet is not an open dataset as its official release consists of pre-computed audio features.
Downloading the original audio tracks can be problematic due to YouTube videos gradually disappearing and usage rights issues.
To provide an alternative benchmark dataset and thus foster SER research, we introduce \textit{FSD50K}, an open dataset containing over 51k audio clips totalling over 100h of audio manually labeled using 200 classes drawn from the AudioSet Ontology.
The audio clips are licensed under Creative Commons licenses, making the dataset freely distributable (including waveforms).
We provide a detailed description of the FSD50K creation process, tailored to the particularities of Freesound data, including challenges encountered and solutions adopted.
We include a comprehensive dataset characterization along with discussion of limitations and key factors to allow its audio-informed usage. 
Finally, we conduct sound event classification experiments to provide baseline systems as well as insight on the main factors to consider when splitting Freesound audio data for SER.
Our goal is to develop a dataset to be widely adopted by the community as a new open benchmark for SER research.
\end{abstract}

\begin{IEEEkeywords}
audio dataset, sound event, recognition, classification, tagging, data collection, environmental sound.
\end{IEEEkeywords}
\vspace{-1mm}

% \fi
\IEEEpeerreviewmaketitle

\section{Introduction}
\label{sec:intro}
\input{S1_Intro}

\section{Related Work}
\label{sec:related}
\input{S2_Related_short}

\section{Dataset Creation}
\label{sec:creation}

\input{S3_Creation_short}

\section{Dataset Description}
\label{sec:description}
\input{S4_Description_short}

\section{Experiments}
\label{sec:experiments}
\input{S5_Experiments}

\vspace{-1mm}
\section{Summary and Conclusion}
\label{sec:conclusion}

\input{S6_Conclusion}
\appendices

\input{S9_Appendices}

\vspace{-2mm}
\section*{Acknowledgment}
The authors would like to thank everyone who contributed to FSD50K with annotations, and especially Mercedes Collado, Ceren Can, Rachit Gupta, Javier Arredondo, Gary Avendano and Sara Fernandez for their commitment and perseverance.
The authors would also like to thank Daniel P.W. Ellis and Manoj Plakal of Google Research for valuable discussions.
This work is partially supported by the European Union's Horizon 2020 research and innovation programme under grant agreement No 688382 AudioCommons, and two Google Faculty Research Awards 2017 and 2018, and the Maria de Maeztu Units of Excellence Programme (MDM-2015-0502).
The authors thank Nvidia for the donated GPUs.

\newpage
\ifCLASSOPTIONcaptionsoff
  \newpage
\fi



%\bibliographystyle{IEEEtran}
%\bibliographystyle{IEEEtran}
%\bibliography{IEEEexample}

\bibliographystyle{IEEEtran}
\bibliography{bibliography}

%\bibliography{refs}

%\begin{thebibliography}{1}


%\end{thebibliography}


\begin{IEEEbiography}[{\includegraphics[width=1in,height=1.25in,clip,keepaspectratio]{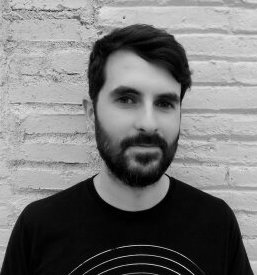}}]{Eduardo Fonseca}
is a researcher with the Music Technology Group, Universitat Pompeu Fabra, where he received his Ph.D. degree under the supervision of Dr. Xavier Serra, partially funded by two Google Faculty Research Awards. His research interests include audio dataset creation and learning algorithms for sound event recognition using different types of supervision, including learning with noisy labels and self-supervision. He received two M.Sc. degrees in Acoustics Engineering from Aalborg University and in Telecommunications Engineering from the Technical University of Madrid. He has interned at Google Research in 2019 and 2020 and his work received one of the Best Paper Awards at WASPAA21. He is actively involved in the DCASE community, where he has served as a Challenge Task Organizer and Technical Program Co-Chair.
\end{IEEEbiography}

\begin{IEEEbiography}[{\includegraphics[width=1in,height=1.25in,clip,keepaspectratio]{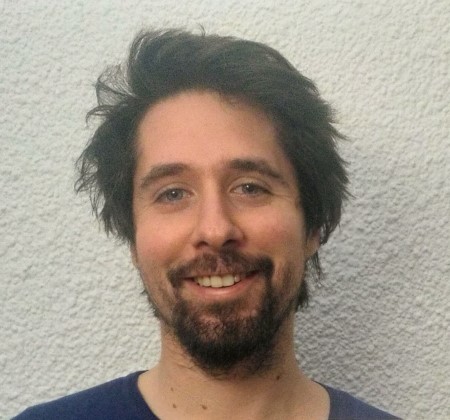}}]{Xavier Favory}
graduated from Ecole Nationale Supérieure de l'Electronique et de ses Applications (France) with an Electronic Engineering Master Degree, specializing in Multimedia Systems in 2014. He then obtained the Master’s degree in Acoustics, Signal Processing and Informatics applied to Music from IRCAM, University Pierre and Marie Curie and Télécom ParisTech in 2015. He earned his Ph.D. in 2021, at the Music Technology Group, Unversitat Pompeu Fabra, where he is now continuing as a post-doctoral researcher. His research interests lie in the areas of audio signal processing, machine learning and human-computer interaction. He is author and co-author of several peer-reviewed publications, and he has been contributing as a reviewer for multiple conferences or workshops (e.g., International Society for Music Information Retrieval, Web Audio Conference, Digital Audio Effects Conference, International Joint Conference on Neural Networks). He also has strong interests in web technologies and development. 
\end{IEEEbiography}

\begin{IEEEbiography}[{\includegraphics[width=1in,height=1.25in,clip,keepaspectratio]{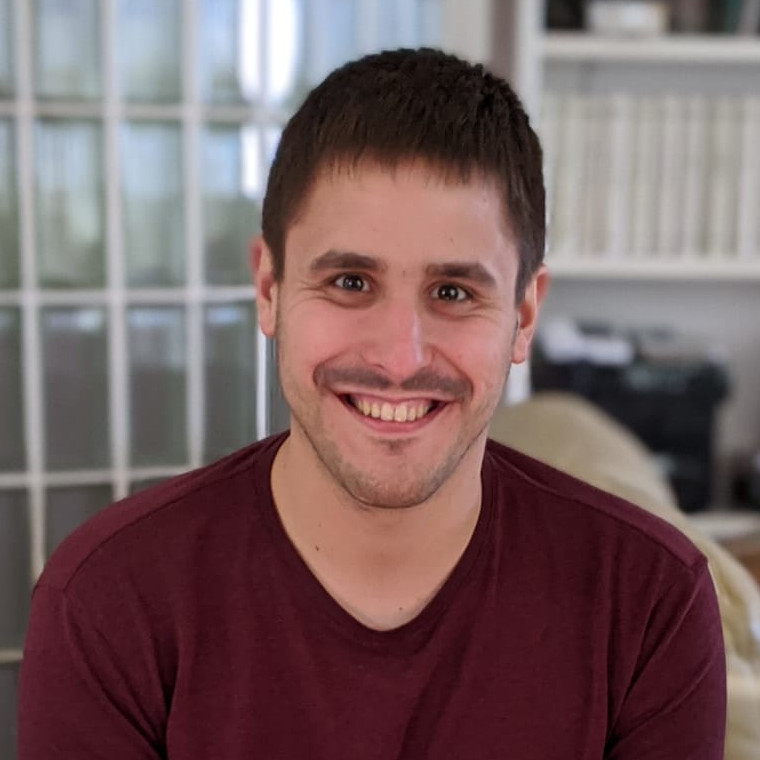}}]{Jordi Pons}
is a researcher at Dolby Laboratories. Previously, he received a PhD in music technology, large-scale audio collections, and deep learning at the Music Technology Group (Universitat Pompeu Fabra, Barcelona). He also recieved a MSc in sound and music computing (Universitat Pompeu Fabra, Barcelona), and his BSc was in telecommunications engineering (Universitat Politècnica de Catalunya, Barcelona). He also interned at IRCAM (Paris), at the German Hearing Center (Hannover), at Pandora Radio (USA, Bay Area), and at Telefónica Research (Barcelona).
\end{IEEEbiography}

\begin{IEEEbiography}[{\includegraphics[width=1in,height=1.25in,clip,keepaspectratio]{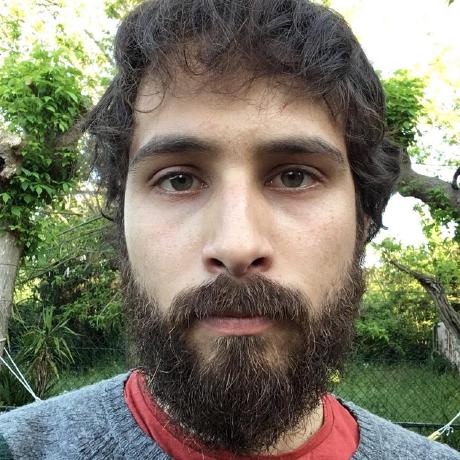}}]{Frederic Font}
is a senior researcher at the Music Technology Group of the Department of Information and Communication Technologies of Universitat Pompeu Fabra, Barcelona, Spain. He received the MSc and PhD degrees in Sound and Music Computing in 2010 and 2015, respectively, from Universitat Pompeu Fabra. His current research is focused on the understanding and analysis of large audio collections, including sound characterisation and classification, to improve sound retrieval techniques and to facilitate the reuse of large audio collections in creative and scientific contexts. Frederic is the coordinator of the Freesound website and related research and development projects, and has recently coordinated the EU funded Audio Commons Initiative.
\end{IEEEbiography}

\begin{IEEEbiography}[{\includegraphics[width=1in,height=1.25in,clip,keepaspectratio]{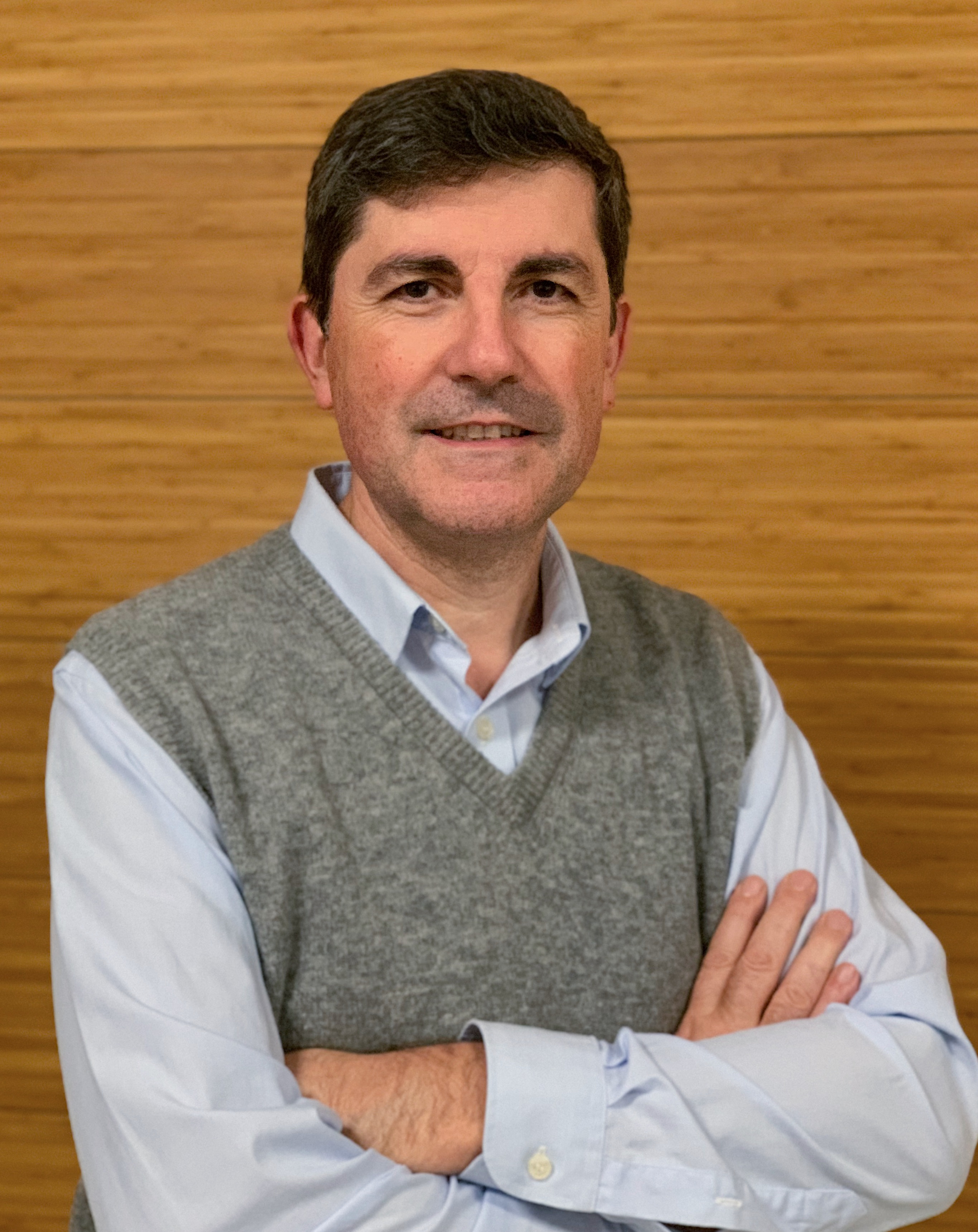}}]{Xavier Serra}
is a Professor of the Department of Information and Communication Technologies and Director of the Music Technology Group at the Universitat Pompeu Fabra in Barcelona. After a multidisciplinary academic education, he obtained a PhD in Computer Music from Stanford University in 1989 with a dissertation on the spectral processing of musical sounds that is considered a key reference in the field. His research interests cover the computational analysis, description, and synthesis of sound and music signals, with a balance between basic and applied research and approaches from both scientific/technological and humanistic/artistic disciplines. Dr. Serra is very active in the fields of Audio Signal Processing, Sound and Music Computing, Music Information Retrieval and Computational Musicology at the local and international levels, being involved in the editorial board of a number of journals and conferences and giving lectures on current and future challenges of these fields. He was awarded an Advanced Grant from the European Research Council to carry out the project CompMusic aimed at promoting multicultural approaches in music information research.
\end{IEEEbiography}


%\newpage



%\vfill


\end{document}

%% file: S1_Intro.tex
%\IEEEPARstart{T}{his} demo file is intended to serve as a ``starter file''for IEEE journal papers produced under \LaTeX\ using IEEEtran.cls version 1.8b and later.

% SER applications / DCASE / 
\IEEEPARstart{S}{ound} event recognition (SER) is the task of automatically identifying the sounds occurring in our daily lives, assigning a label within a target set of sound classes.\footnote{We shall use the expression sound event \textit{recognition} broadly to encompass both sound event \textit{classification} or \textit{tagging} (SET)---a task requiring to identify what sound event classes are present in an audio clip, regardless of start time and end time---as well as sound event \textit{detection} (SED)---a task requiring to localize and identify sound events in an audio clip with start and end times.\label{foot_SERSETSED}}
SER has gained increasing attention in the past few years, becoming a key component in applications related to healthcare \cite{drugman2020audio,huwel2020hearing,messner2020multi}, urban sound planning \cite{bello2019sonyc}, bioacoustics monitoring \cite{cramer2020chirping,lostanlen2019robust,xu2017north}, multimedia event detection \cite{wang2016audio}, large-scale event discovery \cite{jansen2017large}, surveillance \cite{crocco2016audio,sanchez2017maximum}, or noise monitoring for industrial applications \cite{morrison2019otomechanic}.
The SER research community has grown substantially over the last decade, as evidenced by the increasing traction of the \textit{Detection and Classification of Acoustic Scenes and Events} (DCASE) Challenge and Workshop \cite{mesaros2017detection}, which promote research and evaluation on common publicly available datasets.
% History
Early stage works in SER relied on feature engineering approaches using standard machine learning classifiers such as support  vector  machines \cite{foggia2015reliable}, Gaussian  mixture  models \cite{mesaros2010acoustic} or matrix factorization techniques \cite{cotton2011spectral}.
This initial trend was followed by the rapid adoption of deep learning approaches using fully connected neural networks \cite{cakir2015polyphonic}, convolutional neural networks (CNN) \cite{piczak2015environmental}, recurrent neural networks (RNN) \cite{parascandolo2016recurrent}, or combinations thereof \cite{cakir2017convolutional}.
To allow successful exploitation of the data hungry deep learning approaches, it became evident the need for new, larger, and more comprehensive data resources for development and evaluation of SER models.
In contrast, previous SER datasets were of a more limited size and coverage (e.g. \cite{foster2015chime,piczak2015esc,Mesaros2016_EUSIPCO}).
In the current paradigm, datasets in SER are crucial, similarly as in computer vision \cite{sun2017revisiting}, as exemplified by the significant breakthroughs that ImageNet has allowed for image recognition \cite{russakovsky2015imagenet}.
%old In this way, data in SER became essential, similarly as in computer vision \cite{sun2017revisiting}, where, for example, the release of ImageNet allowed significant breakthroughs in image recognition \cite{russakovsky2015imagenet}.

%AudioSet and its problems
To address the lack of large datasets in SER, AudioSet was released in 2017.
AudioSet consists of $\approx$2.1M audio clips manually labeled using 527 classes \cite{gemmeke2017audio}.
Its unprecedented size, coverage and diversity represented a milestone that has transformed SER research.
However, in our view, AudioSet has the major shortcoming of not being an open dataset, as we explain next.
Specifically, AudioSet is composed of audio tracks taken from YouTube videos, which are not freely distributable due to YouTube Terms of Service.
This is the reason why AudioSet is released as a dataset of audio features (instead of audio waveforms),\footnote{\url{https://research.google.com/audioset/download.html}\label{foot_audioset_download}} which are extracted at a time resolution of 960ms using a pre-trained model.
This limits the adoption and flexibility of a number of SER methods.
For this reason, some researchers opt to download and use the audio tracks from the original YouTube videos, despite the intrinsic issues entailed in this process.
These issues include the burden of downloading a massive amount of data from a non-official release, and the fact that the constituent videos are gradually disappearing.
More specifically, videos can become unavailable due to a variety of reasons such as deletions of videos or user accounts, privacy issues, copyright claims, or country-dependant availability.
In an attempt to download the AudioSet audio tracks, we could download 18,205 from 20,371 evaluation segments, and 19,862 from 22,160 balanced train segments---a loss of 10.6\% and 10.4\% respectively.\footnote{Data from May 11th, 2020.}
The fact that the amount of evaluation and train clips available decreases over time with non-negligible differences limits AudioSet suitability for systems' benchmarking.

% not much after AudioSet================ % From AudioSet til today & GAP
After AudioSet, several efforts in dataset creation for SER have been made (e.g. \cite{cartwright2019sonyc,turpault2019sound,Adavanne2019_DCASE,gharib2019voice,Dekkers2017,Fonseca2019learning,Fonseca2019audio,cartwright2020sonyc}).
Nonetheless, these recent datasets are task/domain-specific, or of a much more limited coverage (e.g., usually featuring few tens of classes), and some of them are composed of synthetic audio material.
This contrasts with the computer vision field, where major efforts have been made to collect large and general-purpose datasets as alternatives to ImageNet (e.g. \cite{lin2014microsoft,li2017webvision,kuznetsova2018open}), allowing benchmarking on complementary recognition problems.
Thus, the SER field lags far behind in terms of dataset availability, and we believe that open and sustainable dataset creation initiatives are needed to foster SER research and, more generally, machine listening research.
In addition, we think it is important to document at length the main aspects of data collection and curation when releasing a dataset---a common practice in computer vision \cite{russakovsky2015imagenet,kuznetsova2018open} that has also recently been proposed in audio research \cite{mcfee2018open}.
Making this information available allows researchers to incorporate data-informed decisions in the design of learning pipelines and in the analysis of results, and can also serve as inspiration for potential dataset creators.

% In this paper…
To address these issues and foster SER resesarch, in this paper we introduce \textit{FSD50K} (\textbf{F}ree\textbf{s}ound \textbf{D}ataset \textbf{50k}): a dataset containing 51,197 audio clips totalling over 100h of audio manually labeled using 200 classes drawn from the AudioSet Ontology.
The audio clips are gathered from Freesound and are licensed under Creative Commons (CC) licenses, which allow easy sharing and reuse, thereby making the dataset freely distributable (including audio waveforms).
To our knowledge, this is the largest fully-open dataset of human-labeled sound events, and the second largest after AudioSet.

\vspace{-2mm}
\subsection{Contributions}
%\IEEEpubidadjcol
Our contributions are as follows:
\begin{enumerate}
    \item a human-labeled open dataset primarily designed for the development and evaluation of multi-label sound event classification systems, but that also allows a variety of sound event research tasks,
    \item a detailed description of the FSD50K creation process tailored to the particularities of Freesound data, including challenges encountered and solutions adopted (Sec. \ref{sec:creation}),
    \item a comprehensive characterization of the dataset along with discussion of limitations and key factors to allow its audio-informed usage (Sec. \ref{sec:description}), and
    \item a set of sound event classification experiments to provide baseline systems as well as insight on the main factors to consider when splitting Freesound audio data for SER tasks (Sec. \ref{sec:experiments}).
\end{enumerate}

The information here presented is useful to researchers using FSD50K (and in general using Freesound data for machine learning) as it allows making data-informed decisions for design choices of machine listening systems.
It may also be useful for researchers working on the creation of large-vocabulary datasets.
In addition to the audio waveforms and ground truth, FSD50K includes metadata used during the creation process as well as Freesound metadata for the clips forming the dataset (Sec. \ref{ssec:characteristics}).
All these resources can be downloaded from Zenodo.\footnote{\url{https://doi.org/10.5281/zenodo.4060432}\label{foot_zenodo}}
Likewise, code\footnote{\url{https://github.com/edufonseca/FSD50K_baseline}\label{foot_github}} for baseline experiments and a companion site\footnote{\url{https://annotator.freesound.org/fsd/release/FSD50K/}\label{foot_companion}} for FSD50K is also available.
The companion site allows exploring the audio content of FSD50K as well as reporting labelling errors.

%% file: S2_Related_short.tex
This Section discusses the most important datasets for sound event tagging (SET),\textsuperscript{\ref{foot_SERSETSED}} listed in Table \ref{tab:dataset_sota}.
% The analogous for sound event detection (SED) datasets can be found in Appendix \ref{app:sed_datasets}. 
The datasets are selected based on number of Google Scholar citations, as well as popularity and/or size for the most recent ones. 
% Table  summarizes some aspects of a few most relevant SET datasets. 
For comparison, the proposed FSD50K is listed at the bottom.
%of a smaller selection of the few most relevant datasets, in our view, after discarding synthetic datasets. 
%\textit{m-c} and \textit{m-l} in the \textit{task} column correspond to multi-class and multi-label.
The basic common aspect in SET datasets is that labels are provided at the clip-level (without timestamps), usually regarded as \textit{weak} labels.
This contrasts with sound event detection (SED) datasets, where sound events are labeled using also start and end times (usually regarded as \textit{strong} labels).
We denote \textit{multi-class} datasets (\textit{m-c}) as datasets where each example is labeled with only one class label, which are sometimes denoted as \textit{single-label datasets}.
In a \textit{multi-label} dataset (\textit{m-l}), in contrast, each example can be labeled with \textit{one or more} class labels.
\begin{table*}[h!]
% \begin{minipage}{\textwidth} % <--- new
\vspace{-2mm}
\caption{A selection of most relevant datasets for SET. \textit{m-c} and \textit{m-l} correspond to multi-class and multi-label}
\vspace{-2mm}
%\vspace{+3mm}
\centering
\begin{tabular}{l|ccccccc}
\toprule
\textbf{dataset}             & \textbf{clips} & \textbf{clip length} & \textbf{duration}  & \textbf{classes} & \textbf{task} & \textbf{source} & \textbf{domain/task}\\
\midrule
UrbanSound8K \cite{salamon2014dataset} (2014)  & 8732         & $\leq$4s        &8.8h        &10 bal    & m-c   & Freesound  & urban sounds \\
ESC-50 \cite{piczak2015esc} (2015)            &2000           &5s               &2.8h        &50 bal    & m-c  & Freesound   \\ 
CHiME-home \cite{foster2015chime} (2015)      & 6138          & 4s              &6.8h        &7 unbal   & m-l   & CHiME\footnotemark    & domestic sounds\\ 
%TUT Sound events 2016 \cite{Mesaros2016_EUSIPCO} & 32   & 3-5min    & $\approx$2h   &18 unbal   & m-l SED   & TUT  \\
%TUT Sound events 2017 \cite{DCASE2017challenge}  & 32   & 3-5min    & $\approx$2h   &6 unbal   & m-l SED   & TUT  \\
AudioSet \cite{gemmeke2017audio} (2017)    & $\approx$2.1M   &$\approx$10s    &$\approx$5731h &527 (un)bal  & m-l  & YouTube  \\ 
%SINS (2018)  \cite{Dekkers2017}               & xxx        & 10s            & xxx          &9   & XXXXX      & SED, m-c, unbalanced domestic sounds \\
FSDnoisy18k \cite{Fonseca2019learning} (2019) & 18,532    & 0.3-30s         & 43h         & 20 unbal    & m-c   & Freesound  & noisy labels\\ 
FSDKaggle2019 \cite{Fonseca2019audio} (2019)   & 29,266   & 0.3-30s         & 103h          & 80 unbal & m-l   & Freesound/YFCC100M\footnotemark  & noisy labels \& domain mismatch\\ 
SONYC-UST-V2 \cite{cartwright2020sonyc} (2020) & 18,510     & 10s           & 51h          & 31 unbal & m-l   & SONYC\footnotemark  & urban sounds\\ 
\textbf{FSD50K} (2020)                         & 51,197     & 0.3-30s         & 108h         & 200 unbal & m-l   & Freesound \\ 

%\midrule
\bottomrule
\end{tabular}
\label{tab:dataset_sota}
\vspace{-3mm}
% \end{minipage}
\end{table*}
\footnotetext[7]{\url{http://spandh.dcs.shef.ac.uk//projects/chime/}}
\footnotetext[8]{\url{https://multimediacommons.wordpress.com/yfcc100m-core-dataset/}}
\footnotetext[9]{\url{https://wp.nyu.edu/sonyc/}}
%-----------Before the release of AudioSet, the most widely used datasets were XYZ,
\vspace{-2mm}
\subsection{Datasets Released Before AudioSet}
Before the release of AudioSet, the most widely used datasets for SET have been UrbanSound8K \cite{salamon2014dataset}, ESC-50 \cite{piczak2015esc}, and to a lesser extent CHiME-home \cite{foster2015chime}.
All of them feature short audio chunks and a total duration of less than 10h. 
Curiously, the two former are one of the few multi-class balanced datasets in SET---most datasets are unbalanced and/or multi-label---and also the most widely used (besides AudioSet).
UrbanSound8K and CHiME-home count with a significant amount of clips per class; nonetheless, part of this abundance comes from the fact that many clips are actually time slices coming from the same original recording.
For example, UrbanSound8K is sourced from 1302 Freesound clips.
ESC-50 features a large vocabulary (50 classes) when compared to other datasets from 2014/2015, but it suffers from data scarcity (only 40 clips/class).
Common to all mentioned datasets is that they provide a k-fold cross validation setup---a practice that tended to disappear after the AudioSet release.

\vspace{-2mm}
%\subsubsection{AudioSet and AudioSet Ontology}
\subsection{AudioSet}
Google's AudioSet is the largest dataset of sound events released to date, consisting of $\approx$2.1M audio clips manually labeled using 527 classes of the AudioSet Ontology \cite{gemmeke2017audio}.
AudioSet is the first dataset to put emphasis on general-purpose SER, enabling sound event recognizers to describe a large variety of sound classes, thus aiming at the transcription of most everyday sounds.
AudioSet is split into a train and an evaluation set, and it is highly imbalanced, with some classes being particularly common (e.g. \textit{Music} and \textit{Speech}) while others are much more scarce (e.g. \textit{Toothbrush}).
The public release provides a balanced train partition of 22,176 clips in addition to the full unbalanced train set.
While the dataset is manually labeled in full, 
%(which entails a tremendous endeavour)
its unprecedented size and coverage comes at the expense of a less precise labeling.
The amount of labeling error is estimated at above 50\% for $\approx$18\% of the classes.\footnote{See \url{https://research.google.com/audioset/dataset/index.html} for details on how the quality is estimated, accessed 25th June 2020.\label{foot_aset_noise}}
%As mentioned in Sec. \ref{sec:intro}, the official release provides audio features instead of audio waveforms.
Recently, strong labels for a small portion of AudioSet ($\approx$81k clips) were released \cite{hershey2021benefit}.
% The AudioSet Ontology (a subset of which is used to organize FSD50K) is described in Sec. \ref{ssec:acqui}.

\subsection{Datasets Released After AudioSet}
After AudioSet, some of the released datasets for SET are task-dependent, designed to enable the study of particular problems. 
Examples include FSDnoisy18k \cite{Fonseca2019learning} or FSDKaggle2019 \cite{Fonseca2019audio}, focused on learning in conditions of noisy labels and/or acoustic mismatch.
Other datasets are domain-specific, with a vocabulary focused on a specific scope, such as SONYC-UST-V2 for urban sounds \cite{cartwright2020sonyc}.
Compared to pre-AudioSet datasets, these are slightly larger, especially in terms of duration as they feature longer clips (sometimes of variable length), but also in terms of vocabulary.
In addition, they are unbalanced, and the default data split transitioned to a development/evaluation (or train/test) separation.
% Beyond those listed in Table \ref{tab:dataset_sota}, another large dataset is BirdVox-14SD for bird sounds \cite{cramer2020chirping}.
%Beyond those listed in Table \ref{tab:dataset_sota}, another large dataset is BirdVox-14SD, totalling over 14k clips of avian flight calls with 18 species annotation \cite{cramer2020chirping}.
Lastly, a recent large-vocabulary dataset with a substantial amount of data is VGGSound \cite{chen2020vggsound}, an audio-visual dataset consisting of $\approx$200k video clips from YouTube encompassing 300 classes.
However, VGGSound presents several shortcomings for SET.
The focus is put on audio-visual correspondence since the dataset is created mostly through automatic computer vision techniques---hence some classes have a clear visual connotation, e.g., \textit{people eating noodle}.
Also, while the dataset is singly-labeled (one machine-generated label per clip), the authors recognize that clips can contain a mixture of sounds.
% Upon inspection of the VGGSound vocabulary, it seems likely that sound events from different classes co-occur in the same clip (of 10s length), thus creating missing labels---for example, \textit{cat growling} and \textit{cat meowing}, or a combination of \textit{sea waves}, \textit{sailing}, or \textit{wind noise}.
%, or \textit{ocean burbling}.
Missing labels are a form of label noise found to impact sound recognizers \cite{fonseca2020addressing}. 
While measures can be taken to mitigate their effect on training, in evaluation they can lead to misleading results---an issue that we specifically address in FSD50K (Sec. \ref{ssec:task_gen}).
In addition, VGGSound suffers from the intrinsic problems of being based on YouTube (Sec. \ref{sec:intro}).

To our knowledge, all datasets in Table \ref{tab:dataset_sota} are labelled manually (except a portion of FSDnoisy18k and FSDKaggle2019 purposely included for the study of noisy labels).
%All listed dataset releases are freely available including waveforms (except AudioSet).
FSD50K is a superset of the human-labeled portions of FSDnoisy18k and FSDKaggle2019, except for a few audio clips that have been discarded during FSD50K’s curation process.

%% file: S3_Creation_short.tex
\subsection{Design Criteria}
\label{ssec:design}
%(0.5)
% this belongs to \subsection{Overall procedure} but it is here for latex drama
\begin{figure*}[ht]
%\begin{figure}[ht]
    \vspace{-3mm}
    \centering
  \centerline{\includegraphics[width=0.95\textwidth]{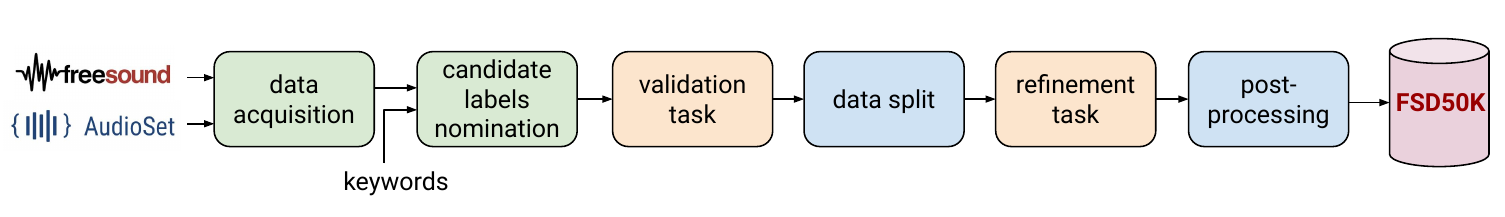}}
    \vspace{-3mm}
    \caption{Overall process of the creation of FSD50K. The process starts from Freesound and the AudioSet Ontology. Stages in green involve automatic data mining, stages in orange correspond to manual annotation tasks, and stages in blue involve data processing to shape the dataset.}
    \label{fig:overall}
    \vspace{-3mm}
\end{figure*}
% font size mas peque, y cajas tb / fix logos / mirar como quedan los colores en IEEE SPL
As design criteria, we set three basic goals and another three specific goals.
The basic goals are:
\textit{i)} the dataset must be open and fully distributable, \textit{ii)} it must contain a large vocabulary of everyday sounds, and \textit{iii)} it must be expandable in terms of data and vocabulary.
To fulfil these basic goals, we turn to Freesound as a source of data, and to the AudioSet Ontology as a vocabulary to organize the data.
Not only do these resources feature a large amount of data and classes, respectively, but Freesound is constantly growing through user uploads, and the ontology is large and was designed to be expandable.
% , allowing dataset expansions. 
% They are described in Sec. \ref{ssec:acqui}.
In addition, we set three specific goals related to the labeling of the dataset and to the emphasis put on the evaluation set.

\subsubsection{Weak Labels}
% We opt to label the dataset with weak labels.
%, that is, clip-level labels expressing the presence of a sound event, without temporal location.
The motivation to label FSD50K with weak labels (i.e., at the clip-level) is that gathering them is simpler, less time consuming and less ambiguous than determining events’ onset/offset (i.e., strong labels).
Weakly supervised learning has demonstrated effectiveness to learn sound event recognizers, both for classification and detection \cite{mcfee2018adaptive}. 
Nonetheless, using weak labels imply certain limitations on training and evaluation, which we highlight in Sec. \ref{ssec:discussion}.

\subsubsection{Label Quality and Dataset Size}
The sound event recognition (SER) field has witnessed a transition away from small and exhaustively labeled datasets (e.g., \cite{piczak2015esc,salamon2014dataset,foster2015chime}), in favour of larger datasets that inevitably include less precise labelling, such as AudioSet \cite{gemmeke2017audio}.
This occurs mainly because it is not feasible to exhaustively annotate large amounts of sound event data.
In our case, we want to seek a trade-off by prioritizing label quality while ensuring a certain amount of data.
%, thus aiming at a medium-size dataset with high label reliability. 
Yet, label noise problems also appear in FSD50K, as in any sound event dataset of a certain size (Sec. \ref{ssec:limit}).

\subsubsection{Emphasis on Evaluation Set}
This is perhaps the design criteria that mostly determines the creation of FSD50K.
%One design criteria that especially drives the creation of FSD50K is to put emphasis on the evaluation set.
An evaluation set defines the target behavior in a recognition task, which makes it possibly the most critical part of a dataset.
Consequently, having a comprehensive, diverse, reliably annotated, and real-world representative evaluation set is important for meaningful systems’ benchmarking.
%CV
The importance of reliable evaluation sets is highlighted by recent research in computer vision which focuses on improving the evaluation and/or validation sets of widely-used datasets---\cite{barz2020we} for CIFAR-10/-100 \cite{krizhevsky2009learning}; and \cite{recht2019imagenet,beyer2020we} for ImageNet \cite{deng2009imagenet}.
%several training alternatives  - but carefully labeled eval sets are still needed.
In addition, alternative learning paradigms to the traditional supervised learning (using reliably-labelled datasets) start to be promising nowadays.
In particular, significant progress is being made in the development of sound event recognizers with noisy supervision \cite{fonseca2020addressing,Fonseca2019learning} or self-supervision \cite{fonseca2021unsupervised,fonseca2021selfsupervised,jansen2018unsupervised}.
While these alternatives can minimize the problems of labelling inaccuracies in the development set, or the need for a labeled development set at all, a carefully curated evaluation set is still critical for benchmarking.
%, regardless of the type of training supervision.
Relatedly, abundant data resources for training are already available, either from AudioSet, or directly from web audio repositories such as Freesound or Flickr (provided appropriate learning strategies are used).
By contrast, to our knowledge, large-vocabulary, carefully-curated evaluation benchmarks are rare---the most prominent being AudioSet’s evaluation set, which suffers from issues of label noise, stability and/or openness (Secs. \ref{sec:intro} and \ref{sec:related}).
By prioritizing the curation of the evaluation set, we contribute to fill this gap.

To tackle the task of labelling the dataset, two approaches are considered: \textit{i)} manual annotation and \textit{ii)} semi-automatic methods based on Active Learning (AL).
% manual
Manual annotation is the conventional approach to dataset labelling, as done in AudioSet \cite{gemmeke2017audio} or ImageNet \cite{deng2009imagenet}.
While this option is laborious and time consuming, when done properly, we believe it leads to more reliable results than involving automatic methods in the annotation loop.
%AL
As an alternative, AL aims at maximizing performance with limited labelling budget by selecting the most informative data for the model to learn.
%cite Riccardi Hakkani Tur?. 
Usually, AL is based on an iterative process involving humans in the loop where automatic methods select the samples to annotate.
%Annotated samples are commonly used to train models that in turn help to select a new batch of samples to annotate.
Often, portions of unlabeled data are automatically labelled via propagation of human-provided labels to similar examples, or with semi-supervised learning approaches.
Recent works studying AL for SER \cite{han2016semi,wang2019active,shuyang2018active,zhao2020active}
report reduced annotation effort with good performance which, in principle, makes AL appealing for dataset creation.
However, these works focus on recognition tasks with less than a dozen classes, and most of them deal with single-label classification and use pre-labeled datasets, where the human annotation step is simulated by a simple assignment of the existing ground truth.
Extending the methods to a setting like ours is considered out of the scope of this work (albeit an interesting topic for future research).
% An overview of the applicability of AL for SER is provided in Appendix \ref{app:al}.

%decision
In order to obtain a high-quality labelling, and being aware of the amount of data to annotate and the budget available, we decide to annotate the dataset manually, similarly as done with AudioSet.
While this means a higher human effort, it presents two advantages.
First, manually annotating FSD50K gives us a deeper insight into the data that would not have been gained otherwise.
Second, it allows us to have a greater control of the labels gathered, as well as to specify not only the labels but also an estimate of sound predominance (Sec. \ref{ssec:task_val}).
Obtaining a set of labels as reliable as possible for this first release is a more favorable starting point for potential future expansions, which could rely on (semi-) automatic methods to scale up more efficiently at the expense of label noise.

%============================================================================================
\subsection{Overall procedure}
\label{ssec:overall}
%(0.5)
The overall process of the creation of FSD50K is illustrated in Fig. \ref{fig:overall}, starting from Freesound and the AudioSet Ontology, and ending with FSD50K.
In every intermediate stage, we progressively filter out a quantity of audio clips and classes in the vocabulary.
Each stage is described in the next subsections.

\subsection{Data acquisition}
\label{ssec:acqui}
%(0.5)
The starting point for the creation of FSD50K is an abundant source of audio clips, a vocabulary to annotate them, and an infrastructure where they can be loaded and annotation tasks can be carried out. 
These items correspond to Freesound, AudioSet Ontology and Freesound Annotator respectively.

%=============
\subsubsection{Freesound}
Freesound\footnote{\url{https://freesound.org/}} is an online collaborative audio clip sharing site \cite{font2013freesound}, with more than 10 million registered users and over 500,000 audio clips.\footnote{Data from August 1st, 2021.}
% , and an average of 3400 new clips added every month.
%
Audio clips shared in Freesound cover a wide variety of audio content, from music samples to environmental sounds, human sounds or audio effects, to name a few.
In addition, the users who upload the clips also provide metadata, e.g., a title, several tags (at least three per clip), and textual descriptions.
We use the user-provided tags in the creation of FSD50K (Sec. \ref{ssec:nomina}).
Since Freesound is collaboratively contributed, it is also very heterogeneous in terms of data origin, recording equipment, and acoustic conditions.
% However, quality is prioritized over quantity in terms of audio recording and associated metadata. 
% One of the most popular use cases of Freesound is the exchange of well-recorded audio samples for creative purposes.
All of the content is CC-licensed, which conveniently allows distribution and reuse.
% As we have seen above, 
Several datasets containing Freesound audio have been widely used by the research community \cite{stowell2013open,salamon2014dataset,piczak2015esc,fonseca2018general,Fonseca2019learning,Fonseca2019audio}, showing its usefulness for dataset creation.

%=============
\subsubsection{AudioSet Ontology}
AudioSet Ontology consists of 632 sound event classes arranged in a hierarchy with a maximum depth of 6 levels \cite{gemmeke2017audio}.\footnote{It can be explored at \url{https://research.google.com/audioset/ontology/index.html}. Sometimes we shall refer to the AudioSet Ontology as \textit{the ontology}.}
The set of classes covers a diverse range of everyday sounds, from human and animal sounds, to natural, musical or miscellaneous sounds.
Within these main sound families, the content covered includes several facets.
The predominant classes correspond to sound events produced by physical sound sources, but there are also some generated by sound production mechanisms (e.g., deformation or impact of materials), and other classes that do not correspond to sound events.
% Then, there is a variety of classes that, strictly, do not correspond to sound events, such as acoustic scenes or classes describing attributes of sound.
Each entry in the ontology includes a textual description among other fields.\footnote{\url{https://github.com/audioset/ontology}}
% The ontology is provided as a list of 632  entries,\footnote{\url{https://github.com/audioset/ontology}} each of them including a textual description among other fields.
Note that the AudioSet vocabulary is a subset of 527 classes drawn from the ontology \cite{gemmeke2017audio}.
% , the remaining being blocked or excluded because of being abstract (see \cite{gemmeke2017audio}).
% onto for our vocabulary
We use the ontology because it is the most comprehensive vocabulary of everyday sounds available, which makes it convenient to cover Freesound's heterogeneity.
% In addition, the rapid acceptance of AudioSet as a resource for SER research has made the AudioSet Ontology a \textit{de facto} standard for everyday sound organization.
Yet, upon careful inspection of the ontology, we realize that improvements could be made in order to make it more 
%consistent as a resource for everyday sound vocabulary, and more 
suitable for organization of Freesound.
However, this task is left out of the scope of this work.
%Even though after careful inspection of the ontology we realize that improvements could be made to make it more consistent as a resource for everyday sound vocabulary, and more suitable for organization of Freesound, this task is left out of the scope of this work.
%Yet, upon a careful inspection, we note that improvements could be made in order to make it more consistent as a resource for everyday sound vocabulary, and to make it more suitable for organization of Freesound audio.
%However, this task was found to be far from trivial and considered out of the scope of this work.
For FSD50K, we focus on a subset of the ontology oriented to most common physical sources, and less oriented to ambiguous or less represented classes in common everyday situations.
Appendix \ref{app:onto} clarifies relevant ontology-related nomenclature used in this paper (such as \textit{leaf} or \textit{intermediate} nodes).

%=============
\subsubsection{Freesound Annotator}
Freesound Annotator\footnote{\url{https://annotator.freesound.org/}} (FSA) is a website that allows the collaborative creation of open audio datasets based on Freesound content.
It serves mainly two goals: the management and exploration of datasets, and the creation and verification of annotations.
% Currently, it only hosts FSD50K.
Originally released on 2017 
% as the \textit{Freesound Datasets} platform in our previous work
\cite{Fonseca2017freesound}, Freesound Annotator has been the object of continuous development.
It started by providing basic prototypes for exploring a taxonomy of audio classes and validating automatically generated annotations.
Additional features were incorporated progressively, including annotation tools and quality control mechanisms (see Secs. \ref{ssec:task_val} and \ref{ssec:task_gen}).
Monitoring tools allow inspection of a dataset progress as well as debugging capabilities.
FSA is an open-source project.\footnote{\url{https://github.com/MTG/freesound-datasets/}}
%primarily developed by XF and FF of the authors.
%FSA uses the Django Python Framework and Postgres as a database.
%A database model allows to store, manage and retrieve clips and their annotations.
%A user model allows to keep track of users' contributions.
%The platform allows setting pioritization schemes to certain data.

\vspace{-2mm}
%============================================================================================
\subsection{Candidate Labels Nomination}
\label{ssec:nomina}
%(0.5)
We started building FSD50K by automatically populating the classes of the ontology with a number of candidate audio clips from Freesound.
Candidate clips were selected by matching user-provided tags in Freesound to a set of keywords associated with every class.
% based on a process of tag matching leveraging the user-provided tags of Freesound.
The goal was to automatically compile a list of candidate labels per clip, indicating potential presence of sound events.
The process consisted of two steps.

First, we compiled a list of keywords for almost every class.
These are terms related to the class label that are likely to be provided by Freesound users as tags when describing audio clips.
Suitable keywords were determined by considering class names and descriptions provided in the ontology, and obtaining the most frequent Freesound tags that co-occur with each target class label.
%We also added the Freesound tags after iteratively inspecting Freesound search results obtained when introducing specific query terms related to each class.
After compiling a first version of the per-class keywords, we manually identified a few classes with very low precision due to pathological inclusion of false positives (e.g., in the \textit{Turkey} class many clips were recordings made in the Eurasian country, instead of containing sounds of the large bird).
To minimize this issue, a refinement process was performed by blocking some tags (e.g., ``turkish" or ``Istanbul" for \textit{Turkey}).
%Therefore, the keywords typically include the class label together with additional similar terms used by the Freesound community.
As an example, the keywords for the \textit{Meow} class are: ``meow”, ``meowing”, ``mew", ``miaow", and ``miaou".
%---the main tags used in Freesound to describe the sound event. 

Second, each class was automatically populated with the corresponding Freesound clips.
We use the compiled lists of keywords as a mapping between clips in Freesound and class labels in the ontology.
Thus, for each clip, all user-provided tags are examined and, when a tag matches a keyword, the clip becomes a candidate clip for the dataset, and the corresponding class label is nominated as a candidate label for the clip.
This process was done by using the Freesound API.\footnote{\url{https://freesound.org/docs/api/}}
We employ the Porter Stemming algorithm for term normalisation to make our matching process more robust \cite{porter1980algorithm}.

%numbers
In this way we were able to map more than 300,000 Freesound clips to the AudioSet classes. 
We decided to filter out clips longer than 90s to avoid very large audio clips (this length limit will be further reduced later on, see Sec. \ref{ssec:task_val}).
No other filters were applied at this stage.
This left us with a total of 268,261 clips with an average of 2.62 candidate labels.
%discussion
This label nomination system induces potential errors as it depends on factors such as class ambiguity and, especially, the choices of Freesound users when providing tags.
However, it has the advantage of allowing easy and rapid retrieval for a large variety of classes without training any classifiers.

%out
The outcome of this stage is a list of automatically-generated candidate labels per clip, indicating the potential presence of sound events.
%The pool of candidate labels will be human-validated in the next stage.

\vspace{-3mm}
%============================================================================================
\subsection{Validation Task}
\label{ssec:task_val}
%(1)
The goal of this stage is to manually validate the candidate labels nominated in the previous stage.
%The input to this stage is

\subsubsection{Initial Prototype of the Annotation Tool}
To this end, we designed an annotation tool that was deployed in FSA.
Essentially, human raters are presented with a number of audio clips and, for each clip, they must assess the presence of a given sound class. For each class, the annotation process consists of two phases.
First, a \textbf{training phase} where raters get familiar with the class by looking at its location in the AudioSet Ontology hierarchy, the provided textual description, and representative sound examples.
Then, a \textbf{validation phase}, in which raters are presented with a series of audio clips from that class (up to 72 clips in 6 \textit{pages} of 12 clips) and prompted the question: \textit{Is $<$class$>$ present in the following sounds?}. (Fig. \ref{fig:val_task2} shows 3 clips within a page of the final prototype). In this initial prototype, raters must select among ``Present", ``Not Present", and ``Unsure", similarly as done in \cite{gemmeke2017audio}.
% \begin{enumerate}
%     \item a \textbf{training phase} where raters get familiar with the class by looking at its location in the AudioSet Ontology hierarchy, the provided textual description, and representative sound examples.
%     \item a \textbf{validation phase}, in which raters are presented with a series of audio clips from that class (up to 72 clips in 6 \textit{pages} of 12 clips) and prompted the question: \textit{Is $<$class$>$ present in the following sounds?}. (Fig. \ref{fig:val_task2} shows 3 clips within a page of the final prototype). In this initial prototype, raters must select among ``Present", ``Not Present", and ``Unsure", similarly as done in \cite{gemmeke2017audio}.
% \end{enumerate}
Along with an audio player and its waveform, links to each clip's Freesound page were made available, where the original tags and descriptions could be inspected to aid the process.
%Ordering of clips was randomized in order to avoid possible bias effects due to presentation order.
    
\subsubsection{Internal Quality Assessment (IQA)}
We used the initial prototype to run an Internal Quality Assessment (IQA) with the goal of \textit{i) } assessing the quality of the candidates produced by the nomination system, and \textit{ii)} collecting feedback about the prototype and annotation task for improvements.
The IQA consisted of validating 12 candidates (1 page of clips) for every class, covering all classes available.
It was carried out by 11 subjects that volunteered to participate, who could leave per-class comments through a text box.
The feedback collected in the IQA revealed that the annotation task is complex due to factors such as ambiguity in some class descriptions or the difficulty of annotating sound events with very high inter- and intra-class variation.

\subsubsection{Final Prototype of the Annotation Tool}
\begin{figure}[ht]
%\begin{figure}[ht]
    %\vspace{-4mm}
    \centering
  \centerline{\includegraphics[width=0.50\textwidth]{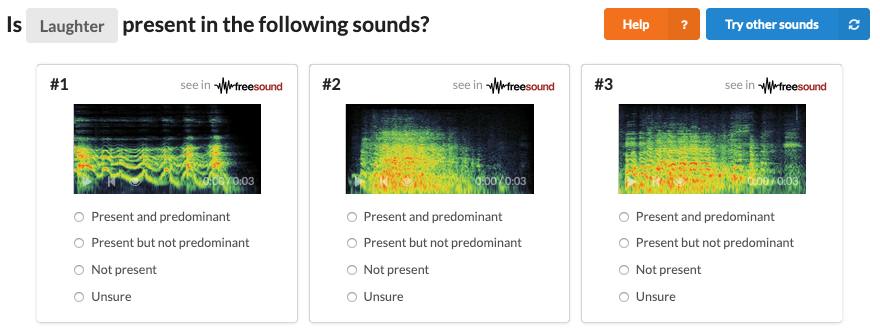}}
  \vspace{-2mm}
    \caption{Screenshot of the ``Validation phase" used for the validation task.}
    \label{fig:val_task2}
    \vspace{-3mm}
\end{figure}

Based on the insight from the IQA, we designed the final annotation tool (Figs. \ref{fig:val_task1} and \ref{fig:val_task2}) which incorporates the following improvements with respect to the previous version:
\vspace{-3mm}
%The fact that this is found difficult by humans reinforces the idea of manually labelling the dataset instead of using semi-automatic methods.
\begin{figure}[ht]
%\begin{figure}[ht]
    %\vspace{-4mm}
    \centering
  \centerline{\includegraphics[width=0.50\textwidth]{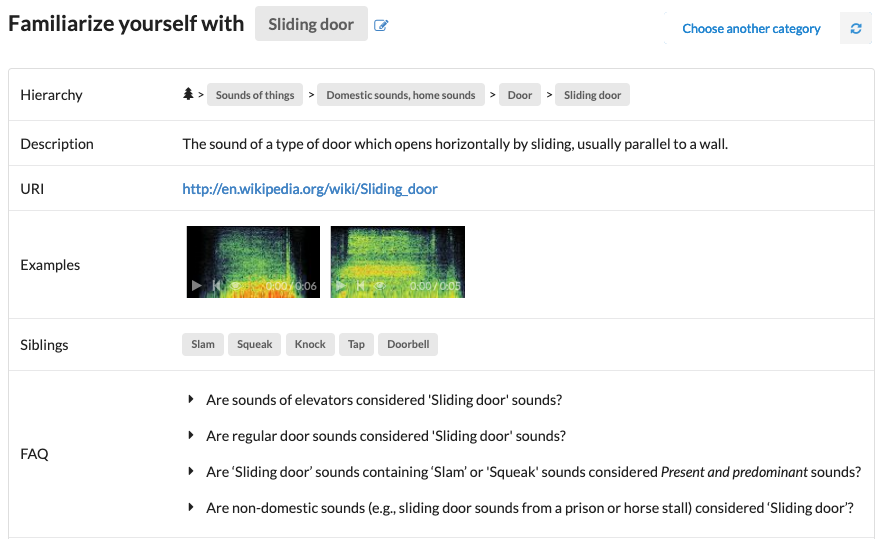}}
  \vspace{-2mm}
    \caption{Screenshot of the ``Training phase" page used for the validation task.}
    \label{fig:val_task1}
    %\vspace{-4mm}
\end{figure}
\vspace{4mm}
\begin{itemize}
    \item Some AudioSet class descriptions were found to be ambiguous, allowing multiple interpretations and generating doubts as to the class scope.
    We decided to include a list of \textbf{Frequently Asked Questions (FAQs)} in each class description to help homogenize raters’ judgment and gather more consistent annotations (see  Fig. \ref{fig:val_task1}). The full FAQ list is provided with the dataset.
    \item In some audio clips, several sound events co-existed with different predominance or salience levels, making the ``Present" response rather ambiguous for raters. To address this issue, we decided to \textbf{split the ``Present" response} into ``Present and predominant" (PP) and ``Present but not predominant" (PNP), as specified in Table \ref{tab:responsetypes}.\footnote{Hereafter, we shall use ``Present" to refer to the union of PP and PNP.} A similar approach was used by Salamon et al. \cite{salamon2014dataset}. The main motivation is to ease the annotation task by mitigating a systematic doubt. As an additional benefit, this distinction allows to separate a subset of clips containing mostly isolated and clean sound events (PP ratings) vs. others featuring events from several classes and/or in more adverse acoustic conditions (PNP ratings). This could allow defining robustness tasks such as training or evaluating with a subset of data of more adverse conditions, similarly as done in \cite{serizel2020sound} for sound event detection (SED) or with ImageNet-A for image recognition \cite{hendrycks2019natural}. Further, the PP/PNP distinction can be useful for source separation studies \cite{Wisdom_InPrep2020}. We note, however, that this distinction is subjective and these ratings should be used as a rough indication.
\begin{table*}[]
\vspace{-3mm}
\centering
\caption{Response types for the validation task}
\vspace{-2mm}
\begin{tabular}{ll}
\toprule
\textbf{Response type}	    &  \textbf{Meaning}   \\ \hline 
Present and 			&The type of sound described is \textbf{clearly present} and \textbf{predominant}.\\
predominant	(PP)			&This means there are no other types of sound, with the exception of low/mild background noise.\\	
\midrule
Present but not 		&The type of sound described is \textbf{present}, but the audio clip also\\
predominant (PNP)			& \textbf{contains other salient types of sound and/or strong background noise}.\\
\midrule
Not Present	(NP)			&The type of sound described is \textbf{not present} in the audio clip. \\
\midrule
Unsure (U)					&\textbf{I am not sure} whether the type of sound described is present or not.\\
\bottomrule
\end{tabular}
\label{tab:responsetypes}
\vspace{-2mm}
\end{table*}
    \item To automatically assess the reliability of the submitted responses, we added \textbf{quality control mechanisms} such as the periodic inclusion of verification clips. Whenever the response for one of these clips is wrong, the responses submitted in a given time span are discarded---a common practice in crowdsourcing platforms.
    \item To further ensure high quality annotations, we decided to require \textbf{inter-annotator agreement}. More specifically, each candidate label is presented to several raters until two different raters agree on a response type. Once an inter-annotator agreement is reached, the label is considered as \textit{ground-truth} and it is no longer presented to other raters. A similar practice is done in \cite{foster2015chime,cartwright2019sonyc}.
    \item To facilitate the localisation and recognition of sound events within the audio clips, we added \textbf{spectrogram visualizations}, thereby easing the annotation task \cite{cartwright2017seeing} (the initial prototype featured less-informative waveforms).
    \item Some audio clips can present highly variable loudness, which can be burdensome for the rater and may affect annotation quality. To mitigate this problem, we \textbf{normalize the loudness} of the sound files following the recommendation EBU R-128 \cite{ebu2011loudness}.
    \item To select which audio clips to present to each rater, we adopt a \textbf{prioritization scheme} that ranks clips according to two criteria: \textit{i)} previously rated label-clip pairs that have not yet reached inter-annotator agreement are prioritized to obtain ground truth labels; \textit{ii)} short clips are prioritized over long ones as shorter clips have a higher label density---considered more informative for learning.
\end{itemize}

Beyond these improvements, we took two additional measures to improve annotation efficiency.
First, the selection of candidates in a number of classes had a very low precision possibly due to sub-optimality of the nomination system.
%(measured as the percentage of ``Present" responses) 
Thus, we decided to discard classes with a rate of ``Not Present" responses above 75\%, as well as classes with very few candidates and others deemed highly ambiguous for annotation. 
This left a total of 395 sound classes (a reduction of $\approx$35\%).
Second, participants reported that the initial duration limit of 90s was too long for human validation, and a potential cause for fatigue.
In addition, the supervision given by weak labels applied to such large lengths is rather vague.
Therefore, we decided to discard clips longer than 30s.
%These measures meant a significant reduction of the least useful audio data. 

\subsubsection{Annotation Campaign}
%what
With the final annotation tool, we launched an annotation campaign to validate the candidate labels at scale.
Given that some classes were found to be much more difficult to annotate than others, we decided to gather annotations using both crowdsourcing and hired raters.
We divided the classes according to an estimated level of difficulty, based on feedback from the IQA.
%(e.g., ambiguity, familiarity to the non-experienced annotator, etc.).
Table \ref{tab:crowd} lists the annotation strategies adopted for each subset of classes.
\begin{table}[t]
%\vspace{-1mm}
\caption{Annotation strategies in the validation task}
\vspace{-2mm}
\centering
\begin{tabular}{@{}cccc@{}}
\toprule
\textbf{class difficulty} & \textbf{example} & \textbf{classes} & \textbf{annotation strategy} \\
\midrule
easy                & \textit{Bark}     & 77           & crowdsourcing  \& hired annotators          \\
medium              & \textit{Piano}    & 100          & crowdsourcing   \& hired annotators            \\
difficult           & \textit{Tearing}     & 218         & hired annotators            \\
%\midrule
\bottomrule
\end{tabular}
\label{tab:crowd}
\vspace{-4mm}
\end{table}
Crowdsourcing consists of gathering validations contributed by any voluntary participant. 
We made the classes of easy and medium difficulty publicly accessible from FSA,\footnote{\url{https://annotator.freesound.org/fsd/annotate/}} which was promoted in Freesound forums and social media.
The most difficult classes, where annotation experience was important to provide reliable responses, were kept private.
They were validated by a pool of hired raters who also complemented the crowdsourcing validations in the rest of the classes.

%whooooo, process
In total, over 350 raters contributed, including voluntary participants, six hired raters, and the first three authors of this paper.
The hired raters were subjects with background in audiovisual engineering, including mostly MSc and PhD students from our group, with self-reported healthy hearing.
We opted for a small pool of raters in order to have more control over the annotation process and to obtain annotations that are as consistent as possible.
We recognize this may induce a certain bias, but we rather have consistent annotations with agreed bias than a certain lack of consistency likely resulting from crowdsourcing annotations for the difficult classes (i.e., label noise).
To this end, the hired raters were trained and closely monitored by the authors, discussing doubts and agreeing on the best course of action. 
The list of FAQs were gradually extended as more insight was obtained.
For consistency, raters were asked to validate groups of related classes (e.g., sibling categories), so they could get familiar with specific sections of the ontology \cite{sabou2014corpus}.
They were instructed to perform the task using high-quality headphones in a quiet environment, and taking periodic breaks to mitigate fatigue.
During the campaign, the hired raters acquired solid expertise on the annotation task and a deep knowledge of the ontology.
Therefore, we consider them experts for this task.

%outcome
The outcome of the annotation campaign was 51,684 clips considered valid for the dataset, that is, with at least one ``Present" label.
All the ``Present” labels amount to 59,981, all of them being the result of inter-annotator agreement, except 3390 which include labels with: \textit{i)} only one PP rating and one PNP rating (and nothing else). This can be considered inter-annotator agreement at the ``Present” level;
%, attributing the different PP/PNP choice to subjective interpretation; 
\textit{ii)} only one PP rating (and nothing else); \textit{iii)} only one PNP rating (and nothing else).
The two latter do not meet our definition of ground truth and could be more prone to errors, but were still considered to slightly increase the amount of data.
It must be noted that the set of labels at this point comes from the validation of candidate labels proposed by a simple nomination system, which ultimately relies on the user-provided Freesound tags.
Hence, it is to be expected that some sound events are not covered by the user-generated tags, or they are not proposed by the nomination system, leading to missing ``Present" labels, a common phenomenon in large sound event datasets \cite{fonseca2020addressing,meire2019impact}.
That is, the resulting pool of audio clips have labels that are mostly correct (estimated at 94.3\% in Sec. \ref{ssec:limit}), albeit potentially incomplete which is problematic in evaluation.
To address this issue, after splitting the data into development and evaluation sets (Sec. \ref{ssec:split}), the latter is refined using another annotation tool (Sec. \ref{ssec:task_gen}).
We define \textit{correct} label as a label accompanying an audio clip that accurately identifies a corresponding sound source in the clip.
We define \textit{complete labels} for an audio clip as the set of labels that identify \textit{all the target} sound sources in the clip as per a predefined vocabulary.

%============================================================================================
\subsection{Data Split}
\label{ssec:split}
%(0.5)
The input to this stage is a pool of 51,684 audio clips with mostly correct labels (albeit potentially incomplete).
The goal is to split the data into two subsets: \textit{development} and \textit{evaluation}.
The development set will be used for training and validation.
The evaluation set will be used for system benchmarking after exhaustive annotation.
As stated in Sec. \ref{ssec:design}, the evaluation set is our priority.
A high quality evaluation set must be comprehensive, varied, and representative \cite{virtanen2018computational}, while being free from contamination from the development set in order to allow testing models’ generalization capabilities.

\subsubsection{Split Criteria}
We set four criteria for the split.

\textbf{Non-divisibility of uploaders.} The issue of \textit{contamination} must be considered when splitting audio data, especially if portions of the data share a common pattern that brings acoustic similarity among its constituents.
In Freesound, audio content is uploaded by users (in the following, \textit{uploaders}).
The uploaders can be very diverse: some are \textit{small}---they upload a small number of audio clips (e.g., up to only few tens)---while other uploaders contribute with hundreds of clips.
In the latter case, it can happen that some of the uploaded clips share the same sound source and/or physical location and/or recording gear (e.g., several notes of the same music instrument or vocalizations of the same pet). 
If some of these recordings are used for training and others for evaluation, their similarity may lead to overly optimistic performance, reflecting the classifier's ability to overfit development examples. 
As a result, this classifier may suffer from performance drop when tested on unseen data.
This issue can be called \textit{weak contamination} between development and evaluation, although, for simplicity, we will refer to it as \textit{contamination} hereafter.\footnote{This should not be confused with \textit{data leakage}, which happens when the \textit{same} (not similar) examples are used for both training and evaluation.}
This phenomenon has been detected in computer vision benchmarks like CIFAR-10 and CIFAR-100 \cite{barz2020we}.
Another example of this in the field of music recognition is the denominated ``album effect” \cite{whitman2001artist,mandel2005song} or ``artist effect” \cite{flexer2009album}.
% Another case of contamination happens when a group of clips captured with the same sensor is split in training and evaluation \cite{cartwright2019sonyc}.
To avoid this issue, we make sure that \textit{all} the content of each uploader is allocated either in the development or evaluation set.
By doing this we assume that the evaluation performance reflects the model’s ability to generalize to new audio material and recording conditions.

%only in long version
%Additional sources of contamination include \textit{i)} audio clip re-purposing (when user $u_a$ downloads a clip by user $u_b$, modifies it, and uploads it as a new clip), and \textit{ii)} the possibility of the same individual creating two user accounts to upload clips.
%While these issues were not considered, based on prior experience, they happen rarely and their effect is considered negligible.

\textbf{Small uploaders for evaluation.} 
%Having defined the uploader non-divisibility constraint, selecting a set that is most suited for evaluation implies several axes.
To obtain a varied evaluation set, it seems reasonable to allocate the content from small uploaders as it guarantees a higher diversity of sound sources, acoustic environments and recording equipment.
In addition, a closer look at the Freesound data distribution revealed that recordings uploaded by small uploaders tend to be slightly longer.
It can therefore be expected that, in general, these longer recordings tend to contain more sound events when compared to shorter clips---a considerable portion of Freesound consists of short clips of few seconds featuring a single event. 
Under this assumption, longer recordings would be more real-world representative (see Sec. \ref{ssec:discussion}).
Also, this is a more interesting content to further annotate exhaustively, and also with timestamps to allow future SED evaluations.
%Finally, another benefit of using small users for evaluation occurs in the hypothetical worst case scenario of one user uploading very similar clips. 
%In this case, it is likely that a competitive model predicting one of them correctly would predict correctly all of them, thus generating a fake performance resolution. By selecting small users we mitigate this problem.

\textbf{A coarse class distribution is enough.} A fine-level split carefully matching a target class distribution is not needed at this point, as during the exhaustive labelling of the evaluation set we expect some classes to grow (Sec. \ref{ssec:task_gen}). This will create an imbalance that will need to be compensated.
% Beyond promoting the allocation of non-divisible small uploaders,

\textbf{Focus on leaf nodes.} Among the classes available at this point, we focus on the subset of 113 leaf nodes with more than 100 clips as they are considered the most important classes.

%-----------
\subsubsection{Split Method}
Given the many constraints, off-the-shelf methods such as random sampling, iterative stratification \cite{sechidis2011stratification} or combinatorial optimization algorithms like knapsack problems \cite{toth1990knapsack,cramer2020chirping} are not well suited.
Therefore, we implement an \textit{ad hoc} approach consisting of iteratively allocating uploaders' content to the evaluation set after sorting them appropriately.
%algo
First, we compute a score per uploader $u$ as:
%\ref{eqn:score_eval}
\begin{equation}
\textrm{score}^u = \textrm{n\_labels}_{\max}^u + \frac{1}{K_u} \sum_{k=1}^{K_u} \textrm{n\_labels}_{c_{k}}^u,
\label{eqn:score_eval}
%\vspace{-1mm}
\end{equation}
where $\textrm{n\_labels}_{\max}^u$ is the maximum number of labels provided by the uploader $u$ in any class, $\textrm{n\_labels}_{c_{k}}^u$ is the number of labels provided by $u$ in the class $c_{k}$, and $K_u$ is the number of classes \textit{touched} by $u$ (i.e., those to which $u$ contributes).
Uploaders are sorted in ascending score order and the content of low-score uploaders is transferred first.
With the first term we prevent uploaders with abundant content concentrated in one specific class, and with the second term preference is given to users with low average number of labels per class for diversity.
% accounting for the scattering of $u$ across classes.
%Upon inspection of the top-ranked uploaders, 
% OLD: We found out that by splitting the target 113 leaf nodes, some content associated with the remaining classes is automatically allocated due to the scattering of uploaders across various classes.
We found out that, by splitting the target 113 leaf nodes, some content corresponding to the remaining classes is automatically allocated due to the uploaders contributing also to them.
This content is deemed sufficient as a fine-level class distribution is not the target at this point.
We then proceed to allocate data to the evaluation set following the process shown in Algorithm \ref{alg:eval_split}.
% \vspace{-1mm}
\begin{algorithm}[h]
% \vspace{-4mm}
\label{alg:eval_split}
\DontPrintSemicolon
\SetAlgoLined
\KwData{Empty evaluation set per-class $E=\left\lbrace e_{c_{i}}=0 \right\rbrace_{i=1}^C$, uploaders ranking $\boldsymbol{u}$}
 %initialization\;
 %\For{pass $n = 1,2, \dotsc  N $}{
    \For{class $c_i \in C$}{
      get current evaluation target $t_{c_{i}}$ \;
      \While{$e_{c_{i}} < t_{c_{i}}$}{
        get next uploader $u$ in ranking $\boldsymbol{u}$ with data  in $c_{i}$\;
        $e_{c_{i}} \leftarrow e_{c_{i}}$ + data from $u$ in $c_{i}$ \;
         \For{class $c_k \in K_u$}{
            $e_{c_{k}} \leftarrow e_{c_{k}}$ + data from $u$ in $c_{k}$ \;
        }
      }
    }
 %}
 \KwResult{A candidate evaluation set}
 \caption{Data allocation to evaluation set}
  \vspace{-4mm}
\end{algorithm}
% \vspace{-3mm}
We traverse the $C=113$ classes starting from the least-represented ones since they have less flexibility for data allocation.
For each class $c_i$, we progressively allocate content from the ranked uploaders $\boldsymbol{u}$ until a target amount of data $t_{c_{i}}$ is reached.
$t_{c_{i}}$ is proportional to the total class label count, and rectified to lie in the range from 50 to 100 labels per class.
By default, the maximum uploader size per class $c_i$ (i.e., the maximum number of clips that one uploader $u$ is allowed to contribute to $c_i$) is set to $0.1 t_{c_{i}}$.
Thanks to the proposed sorting, uploaders in the evaluation set do not reach such a maximum in the majority of classes (they often provide one single clip)---if they do, excess clips are discarded in most cases.
However, due to the high uploader diversity, the maximum uploader size had to be increased in a few exceptions.
%Likewise, some uploaders were considered inadequate for evaluation (e.g., too large in some classes) and were manually assigned to development.
Using the proposed scheme, we processed all the 7229 uploaders and we allocated 2794 of them to the evaluation set, totalling 11,466 clips.

The result is two pools of clips disjoint in terms of uploaders: a candidate development set and a candidate evaluation set. The latter is exhaustively labeled in the refinement task.

\vspace{-3mm}
%============================================================================================
\subsection{Refinement Task}
\label{ssec:task_gen}
%(0.5)
%motivation - goal
As mentioned in Sec. \ref{ssec:task_val}, in some clips, the current label sets could be an underrepresentation of the audio content, biased by the idiosyncrasies of the labeling pipeline.
%Why missing labels are dangerous in eval? Had the current label sets were used for evaluation, in some cases, 
% OLD
% This is problematic in evaluation, as classifiers would be penalized when predicting a correct label that happens to be missing from the ground truth.
% This critical issue would limit the utility of the dataset for system's benchmarking.
% new
When evaluating classifiers' predictions against these incomplete label sets, classifiers can be penalized when predicting a correct label that happens to be missing from the ground truth.
To address this issue, we refine the labels in the evaluation set.
The goal is to obtain an \textit{exhaustive} labelling, that is, a labelling as close as possible to the correct and complete transcription of the audio content (for the considered vocabulary of 395 classes).

\subsubsection{Annotation Tool}
\begin{figure}[ht]
%\begin{figure}[ht]
    \vspace{-1.5mm}
    \centering
  \centerline{\includegraphics[width=0.50\textwidth]{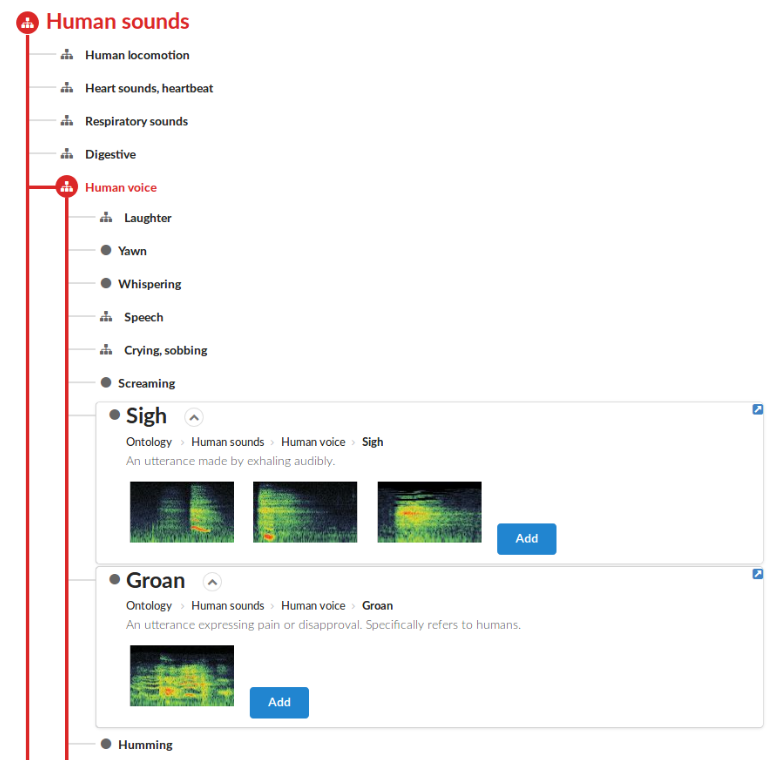}}
      \vspace{-1mm}
    \caption{Table for exploring the ontology in the refinement task.}
    \label{fig:ref_task2}
    \vspace{-3mm}
\end{figure}

We designed and implemented an annotation tool that allows two subtasks:
\textit{i)} to review the existing labels, and \textit{ii)} to add missing ``Present" labels.
%this poses a problem
The subtask of adding missing ``Present" labels has a considerable complexity since audio clips can sometimes contain very different sound events.
Therefore, the success of this task relies on two key factors: \textit{i)} raters with a deep understanding of the ontology, the agreed FAQs, and the particularities of the audio material; \textit{ii)} an interface that facilitates exploration of the large vocabulary of the ontology.
%who
In regard to the first factor we turn to the pool of hired raters (4 of the initial 6), who acquired a solid expertise by extensive participation in the validation task (Sec. \ref{ssec:task_val}).
% task/process
As for the second factor, the refinement task we implemented in FSA includes a tool to interactively explore different depth levels of the ontology (Fig. \ref{fig:ref_task2}). 
This tool is based on a previous version described in \cite{favory2018facilitating}.
A search input box allows to quickly navigate to classes in the table, where their hierarchical context is shown.
For each class, textual descriptions and representative sound examples are displayed.
The interface facilitates the comparison of different classes by simultaneously displaying their information.
%, which was found very useful.

\subsubsection{Annotation Process}
Clips were presented grouped by sound class to facilitate the task. For every class:
\begin{enumerate}
    \item raters were instructed to go through a \textbf{training phase} (same as in the validation task---see Fig. \ref{fig:val_task1}). 
    \item For every clip, they would first \textbf{review existing labels} and modify them if needed. Then, they would \textbf{add any missing labels} by exploring the ontology (Fig. \ref{fig:ref_task2}).
\end{enumerate}
Raters were instructed to provide the most specific labels possible (typically leaf labels) as they are the most informative type of supervision.
The quality assurance practices described for the validation task were also applied in the refinement task.
%Like in the validation task, the raters were closely monitored by the first three authors.
Following this procedure, each evaluation label was verified or reviewed by between two and five different annotators (considering both validation and refinement tasks), including at least one expert.
As a result, labels are expected to be correct and complete in the vast majority of cases (see Sec. \ref{ssec:limit}).
The exhaustive labelling carried out has two implications.
First, absence of labels means absence of sound events (except human error)---a desired feature.
Second, some classes are now much more represented than before as they are prevalent but were underrepresented, thus creating a class imbalance.

The outcome of this stage is a pool of exhaustively labeled clips (for the considered vocabulary), the majority of which will form the evaluation set.

\vspace{-3mm}
%============================================================================================
\subsection{Post-processing}
\label{ssec:postpro}
%(1)
%The goal of this stage is to post-process the incoming pools of data in order to finalize the proposed dataset FSD50K.
This stage starts from two sets of data: a candidate development set with correct but potentially incomplete labels, and an exhaustively-labeled candidate evaluation set.
The vocabulary used so far comprises 395 classes, yet many of them have little data (few tens of clips).
While they may not be adequate for deep learning approaches, they can be useful for other practices, e.g., few shot learning \cite{cheng2019multi}.
Likewise, this information can provide insight as to the specific content of the dataset.
Therefore, we provide two different formats for the annotations in FSD50K:
\textit{i)} The raw outcome of the annotation process, featuring all generated class labels without any restriction. These include classes with few data. We call this the \textit{sound collection} format. \textit{ii)} The outcome of curating the raw annotations into a machine learning dataset with emphasis in sound event recognition tasks. This process involves, mainly, merging low prior classes into their parents thus ensuring a minimum amount of per-class data. This is what we define as \textit{ground truth} for FSD50K, with a vocabulary of 200 classes. 
% \begin{enumerate}
%     \item The raw outcome of the annotation process, featuring all generated class labels without any restriction. These include classes with few data. We call this the sound \textit{collection} format.
%     \item The outcome of curating the raw annotations into a machine learning dataset with emphasis in sound event recognition tasks. This process involves, mainly, merging low prior classes into their parents thus ensuring a minimum amount of per-class data. This is what we define as \textit{ground truth} for FSD50K, with a vocabulary of 200 classes. 
% \end{enumerate}
In Appendix \ref{app:post_pro} we describe the post-processing carried out to obtain what's finally released as FSD50K (consisting of a set of audio clips and corresponding ground truth).
The post-processing stage includes determining FSD50K's vocabulary, balancing development/evaluation sets, creating a validation set, and propagating the labels hierarchically. %to obtain an exhaustive labelling hierarchy-wise.
Further technical details about the sound \textit{collection} format can be found in FSD50K's Zenodo page.\textsuperscript{\ref{foot_zenodo}}

%% file: S4_Description_short.tex
FSD50K is an open dataset of human-labeled sound events containing 51,197 clips unequally distributed in 200 classes drawn from the AudioSet Ontology. 
%This Section discusses its main characteristics, limitations and applications.
The dataset is freely available from Zenodo.\textsuperscript{\ref{foot_zenodo}}
%with both clips and metadata released under CC licenses, 
%and can be downloaded from Zenodo.\textsuperscript{\ref{foot_zenodo}}
Hereafter, we refer to \textit{development} (composed of \textit{training} and \textit{validation}) and \textit{evaluation} sets described in the previous Sections as \textit{dev}, \textit{train}, \textit{val}, and \textit{eval}.

\vspace{-2mm}
%============================================================================================
\subsection{Characteristics}
\label{ssec:characteristics}
%content, classes, leaves nodes
FSD50K is composed mainly of sound events produced by physical sound sources and production mechanisms.
% Hence, the main focus is on the \textit{casual listening} perspective of sound, as defined by Schaeffer \cite{schaeffer2016traite}.
It also includes some classes that can inherently encompass several more specific sources (e.g., \textit{Train}), some classes that do not relate to a specific source but to the perception of sound (e.g., \textit{Clatter}), and few abstract classes (e.g., \textit{Human group actions}).
The dataset has 200 sound classes (144 leaf nodes and 56 intermediate nodes) hierarchically organized with a subset of the AudioSet Ontology \cite{gemmeke2017audio}.
The vocabulary can be inspected in Fig. \ref{fig:perclass}.
Note, however, that in some cases one leaf node in FSD50K (e.g., \textit{Camera}) may be an intermediate node in AudioSet due to the fusion of low-occupancy classes (e.g., \textit{Single-lens reflex camera}) with their parents.
Following AudioSet Ontology's main families, the FSD50K vocabulary encompasses mainly \textit{Human sounds}, \textit{Sounds of things}, \textit{Animal}, \textit{Natural sounds} and \textit{Music}.
%, as well as a few classes belonging to \textit{Onomatopoeia} and \textit{Deformable shell}.
The vast majority of the content corresponds to sounds recorded from a sound field, while a small portion corresponds to sounds captured directly from electronic devices, typically in the context of musical instruments, e.g., some bass drums are generated with drum machines.
% while a small portion corresponds to sounds electronically generated with devices, typically in the context of musical instruments, e.g., some bass drums are generated with drum machines.
%clips, labels, duration
The main characteristics of FSD50K in terms of number of clips, labels, duration and uploaders are listed in Table~\ref{tab:main_stats}.
\begin{table}[ht]
\vspace{-2mm}
\caption{Main statistics for FSD50K}
\vspace{-2mm}
\centering
\begin{tabular}{lccc}
\toprule
             & \textbf{total} & \textbf{dev} & \textbf{eval} \\
\midrule
clips            & 51,197         & 40,966 (80\%)          & 10,231 (20\%)        \\
labels (unpropagated)  & 62,657         & 45,607 (72.8\%)        & 17,050 (27.2\%)      \\
avg labels/clip      & 1.22           & 1.11             & 1.67 \\
labels (propagated) & 152,867        & 114,271              & 38,596       \\
clips w/ leaf label(s)          & 40,461           & 31,310               & 9151      \\
duration         & 108.3h         & 80.4h (74.2\%)          & 27.9h (25.8\%)  \\
avg duration/clip    & 7.6s           & 7.1s             & 9.8s  \\
uploaders          & 7225           & 4936               & 2289      \\
%\midrule
\bottomrule
\end{tabular}
\label{tab:main_stats}
\vspace{-2mm}
\end{table}

\begin{figure*}[t]
%\begin{figure}[ht]
    \vspace{-5mm}
    \centering
  \centerline{\includegraphics[width=0.87\textwidth]{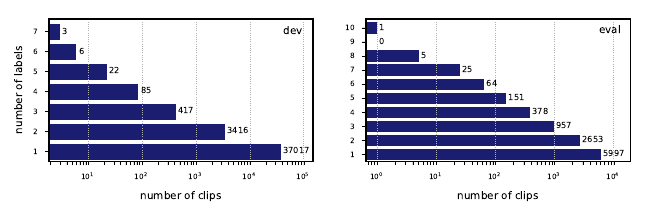}}
    \vspace{-5mm}
    \caption{Label distributions in dev (left) and eval (right) sets. Clips in eval tend to have more labels (by dataset curation). Xaxis scale is logarithmic. Number of labels is reported in the \textit{unpropagated} form. Note that visualization span differs among plots.}
    \label{fig:label_distri}
    \vspace{-3mm}
\end{figure*}
\begin{figure*}[t]
%\begin{figure}[ht]
    %\vspace{-4mm}
    \centering
  \centerline{\includegraphics[width=0.86\textwidth]{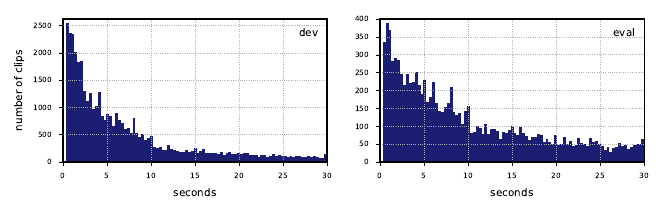}}
    \vspace{-5mm}
    \caption{Audio clip length distributions in dev (left) and eval (right) sets. Clips in eval tend to last slightly longer (by dataset design). Bins correspond to 1/3 second. Note that visualization span differ among plots.}
    \label{fig:clip_len_distri}
    \vspace{-4mm}
\end{figure*}
The audio clips are grouped into a dev split and an eval split such that they do not have clips from the same uploader.
%The dataset is constructed emphasizing the diversity of eval, via inclusion of a higher variety of small uploaders.
Eval is exhaustively labeled, that is, annotations are correct and complete for the considered vocabulary (except human error).
In dev, a small amount of content is exhaustively labeled, but the vast majority is composed of labels that are correct but could be potentially incomplete (see Sec. \ref{ssec:limit} for label noise estimations).
The number of labels is expressed in \textit{unpropagated} and \textit{propagated} forms.
The number of \textit{unpropagated} labels includes only the most specific labels per clip.
It must be noted that this way of counting labels ignores a few labels in cases where a sound event co-occurs with: \textit{i)} events from low prior siblings that were merged with their parent; \textit{ii)} events that do not fit semantically in any other sibling provided by the ontology, hence they are annotated with their parent.
While these cases are not frequent, the true number of human-provided labels describing sound events would be slightly larger than the one reported here.
\textit{Propagated} labels refer to the labels after hierarchical propagation (App.~\ref{app:post_pro}).
We use the unpropagated version to compute the average number of labels per clip.
Note the increased number of labels per clip in eval due to the exhaustive labelling process, as can also be seen by comparing the label distributions in Fig. \ref{fig:label_distri}.
In eval, all classes are present in both the singly-labeled data and the multi-labeled data forming it, except five classes that only appear in the multi-labeled data.
In dev, all classes are present in both the singly-labeled data and the multi-labeled data forming it, except four classes that only appear in the singly-labeled data.
A total of 31,310 clips are labeled with, at least, one leaf label in dev---the remaining 9656 clips are labeled only with intermediate node labels.
This proportion changes significantly in eval, where the majority of clips have leaf labels (9151 out of 10,231)---this is because in the refinement task raters were instructed to provide the most specific labels possible.
% this is due to the label specificity policy used in the refinement task (Sec. \ref{ssec:task_gen}).
All provided ground truth labels are hierarchically propagated, i.e., consistently propagated to their ancestors in the hierarchy.
PP/PNP ratings are provided for the labels validated in the validation task.
Out of the 108.3 hours of human-labeled audio, 31.5 are exhaustively labelled, most of them used for evaluation purposes (eval and val).
The audio clips are of variable length ranging from 0.3 to 30s.
Note the increased average duration of eval clips due to the allocation process (Sec. \ref{ssec:split}), which can also be noticed by comparing the clip length distributions in Fig. \ref{fig:clip_len_distri}.
The ground truth labels are provided at the clip-level (i.e., weak labels).
The dataset is sourced from 7225 Freesound users and the content was uploaded from Freesound's launch in 2005 until early 2019.
%-------end of table

% classes and data
The number of clips per leaf class varies, roughly, from 40 to 200 in eval, and from 50 to 500 in dev, with a few exceptions.
The number of clips in the intermediate nodes grows much more depending on the hierarchy.
Therefore, class imbalance comes from two sources: non-uniform class distribution and variable-length of clips.
%format
The dataset is licensed under CC-BY license---nonetheless, each clip has its own specific license (CC0, CC-BY, CC-BY-NC or CC Sampling+, where CC0 and CC-BY amount to 84.7\% of the dataset).
When original audio clips have more than one channel, they are downmixed to mono.
All clips are provided as uncompressed PCM 16 bit 44.1 kHz mono audio files.
Further details about data licenses, ground truth format, and additional provided metadata can be found in the FSD50K Zenodo page.\textsuperscript{\ref{foot_zenodo}}

%-duration per class. cuales son las clases con clips mas largos. por que? son outdoor? box plot
%-per-class results to show classes

%============================================================================================
\subsection{Discussion}
\label{ssec:discussion}

\subsubsection{Variable Clip Length and Weak Labels}
Labels in FSD50K are provided at the clip-level (i.e., weak labels).
However, unlike other sound event datasets featuring audio clips of same or similar lengths \cite{gemmeke2017audio,salamon2014dataset,salamon2017scaper}, FSD50K is composed of variable-length clips in the range [0.3, 30] seconds (Fig. \ref{fig:clip_len_distri}).
This provides FSD50K with a particular feature.
On the one hand, some clips contain sound events where the acoustic signal fills almost the entirety of the file, which can be understood as strong labels.
To give a sense of this, 12,357 clips in the dev set are shorter than 4s and bear one single label validated only with PP ratings.
Thus, we estimate that the dev set is composed mainly of weakly labeled data and a portion of strongly labeled data, in a rough proportion of 70\%/30\%.
On the other hand, another small portion of the data presents a much weaker supervision---e.g., 9494 dev clips are longer than 10s (Fig. \ref{fig:clip_len_distri}).
The longer the clips, the lesser the certainty of where the labeled event is actually happening. This is referred to as \textit{label density noise} \cite{shah2018closer}, defined as a measure of the weakness of labels for a given weakly labeled clip.
The impact and limitations of weak labels in sound event recognition (SER) are discussed in \cite{turpault2020limitations,Fonseca2019learning}.
% OLD
% In the context of deep networks, clips' variable length implies that audio processing must be done either using fixed-length patches or utilizing variable-length inputs.
% Using fixed-length patches implies two issues: \textit{i)} in training, the weak labels must be inherited by every patch (a practice called \textit{false strong labeling} in \cite{morfi2018data}), which can generate false positives if the label is not active in a given patch;
% \textit{ii)} in evaluation, patch-level scores must be aggregated into clip-level predictions to be compared against the weak labels.
% Utilizing variable-length inputs is free from the issues of the previous approach, but entails certain architectural constraints, such as using fully convolutional networks or appropriate pooling strategies.
Audio processing can be done by handling the variable-length clips \textit{as is}, or by slicing the clips into fixed-length patches. Utilizing variable-length inputs in the context of deep networks entails certain architectural constraints, such as using fully convolutional networks or appropriate pooling strategies. To avoid these constraints, variable-length clips are sometimes split into fixed-length patches, which can be processed in several ways. 
A simple way consists of inheriting the clip-level labels by every constituent patch (a practice called \textit{false strong labeling} in \cite{morfi2018data}), which can generate false positives if the label is not active in a given patch. In this case, for evaluation, patch-level scores must be aggregated into clip-level predictions to be compared against the weak labels.
Another alternative to process fixed-length patches belonging to weakly-labelled clips is to adopt a multiple-instance learning strategy \cite{mcfee2018adaptive}.

\subsubsection{Audio Quality}
Given the heterogeneity of Freesound audio it is difficult to make strong objective claims about audio quality in FSD50K.
Nonetheless, upon inspection of the clips' metadata, it can be seen that many Freesound users utilize (semi-) professional recording equipment (e.g., microphones or preamplifiers of brands such as \textit{Neumann}, \textit{Rode} or \textit{Tascam}).
Our experience after annotating the dataset is that the audio generally has relatively high SNR and dynamic range. 
%of mid to high quality.
To put this into context, we note that the notion of audio quality in sound recognition datasets has changed over time.
In early DCASE Challenges, datasets recorded with professional equipment dominated, some of them being recorded with one single microphone model \cite{mesaros2017detection,Mesaros2016_EUSIPCO,Lafay2017,DCASE2017challenge}.
Then, AudioSet became popular, in which a large variety of devices are used for recording YouTube \textit{videos} (where audio quality is not necessarily a priority), and often including lower SNR conditions.
%After having used and listened to a portion of some of these datasets, we speculate that the overall audio quality of FSD50K lies somewhere between the two aforementioned cases.
We extracted the global SNR for FSD50K and AudioSet (its eval set and balanced train set) using the ITU-T P.563 \cite{malfait2006p}.
It must be noted that P.563 is designed for evaluating human speech, while the content of both datasets is much more diverse. 
Therefore, strong claims cannot be made based on this measurement.
We use it as a common-referenced, rough indication of SNR given that it is not trivial to accurately compute SNR for the different types of audio under consideration.
SNR values reported in Table \ref{tab:fsd_aset} are mean and median of per-clip SNR values.
The mean SNR for FSD50K is greater than that of AudioSet.
For AudioSet, it can also be seen that the mean SNR is greater than the median, suggesting that the SNR distribution is positively skewed, i.e, lower SNR values are more frequent.

\subsubsection{Real-world Audio}
Many clips in Freesound are real-world recordings of sound events happening in the wild, e.g., a car passing by.
However, it is not uncommon that some sound events are recorded under careful conditions in order to obtain clean and isolated high-quality sounds, as in a foley sound setting (e.g., the sound of tearing paper carefully located in front of a microphone).
Further, a few clips in Freesound consist of sound events purposely generated with the sole objective of being recorded, e.g. a faked laughter. 
While these recordings are valuable for sound design, in some cases they could feature a lack of naturalness or acoustic mismatch with respect to sound events in the wild.
This may question the suitability of a portion of the data for learning sound recognizers to be deployed in the wild, where more adverse generation and recording conditions can be encountered.
To what extent this affects models' generalization to adverse scenarios is an open question. Mitigating this potential issue could be a research problem involving, for example, data augmentation \cite{salamon2017deep} or domain adaptation \cite{li2017large} techniques.

\vspace{-2mm}
%============================================================================================
\subsection{Limitations}
\label{ssec:limit}
\subsubsection{Label Noise}
Throughout this paper we have discussed the correctness/completeness of labels in the dataset.
While we aimed at full label correctness and completeness, this is unrealistic as it would mean perfect accuracy of the nomination system that proposes candidate labels (Sec. \ref{ssec:nomina}), and of the human-provided labels.
As supervised learning research moves towards larger datasets, issues of label noise become inevitable.
For instance, labeling error in AudioSet is estimated at above 50\% for $\approx$18\% of the classes.\textsuperscript{\ref{foot_aset_noise}} 
Similarly, ImageNet data are often presumed to have correct labels, but it has been estimated that at least 100k images could be labeled incorrectly \cite{northcutt2019confident}.
In SER, label sets in not-small datasets are inherently noisy due to reasons like sub-optimality of automatic methods used in the creation, or the difficulty of annotating audio---especially without visual cues, with large vocabularies, and because the annotation process is, sometimes, inherently subjective and ambiguous.
% Consequently, recent works have shown the efficacy of label noise treatment in large datasets such as AudioSet \cite{fonseca2020addressing,kumar2019secost} and mid-size datasets \cite{Fonseca2019learning,Fonseca2019model,iqbal2020learning}.

Despite our efforts to mitigate label noise in FSD50K, there are still a few label noise problems.
The main problem is the existence of \textit{missing} “Present” labels (false negatives).
These are labels that would be included in an ideal exhaustive annotation but which are missing from the current set.
Recent work identifies this as a pathology in AudioSet as well, and proposes a method to tackle it \cite{fonseca2020addressing}.
This problem affects the dev set more due to the annotation process based on validation of previously nominated labels---if sound events are not nominated by the system, they lack labels (Sec.~\ref{ssec:task_val}).
This may happen with sound events that tend to be less represented by the Freesound user-provided tags, such as human or bird sounds when they are not the most relevant events in a clip.
Because the eval set received exhaustive annotation, this problem is minimized there.
To a much lesser extent, two additional sources of missing labels exist.
First, the impossibility of propagating labels in the hierarchy when multiple ambiguous paths are encountered---again, this affects more the dev set (see App.~\ref{app:post_pro}).
Second, missing labels can occur as a result of annotating with a finite vocabulary---there may be additional acoustic content \textit{out-of-vocabulary}.
In particular, in the annotation process, when we encounter a sound event out-of-vocabulary, two cases can happen: \textit{i)} if the event is alone in the clip (or with other out-of-vocabulary events), the clip is discarded; \textit{ii)} if the event co-occurs with other events \textit{in-vocabulary}, then the clip is kept, but the out-of-vocabulary event remains unlabeled. 
Apart from missing labels, the other label noise problem is \textit{incorrect} “Present” labels (a false positive, and potentially a false negative if the true class is in-vocabulary).
This is the result of human annotation errors.
Because we adopted mechanisms to bootstrap human annotation quality (Secs. \ref{ssec:task_val} and \ref{ssec:task_gen}), we expect incorrect labels to be rare (see below).
Both missing and incorrect labels would be class-conditional as some classes are clearly more ambiguous than others.
% When labelling errors occur, the non-existent true labels can be either in-vocabulary or out-of-vocabulary, which pose different problems.
More details about label noise characterization can be found in \cite{Fonseca2019learning}.

% quantification of label noise
We can use the refinement task processing to quantify the label noise at the output of the validation task.
A total of 11,847 audio clips (carrying a total of 13,681 labels) were processed with the refinement task, the majority of which ended up in the eval set.
The processing undergone by these clips in this task in order to approach a complete sound event transcript is summarized next, in order to quantify the label noise at this point.
First, we estimate the amount of missing labels.
A total of 6030 clips (50.9\%) received at least one additional label, indicating that there was some unlabeled material. For these 6030 clips, a total of 10,473 labels were generated. 
Second, we estimate the amount of incorrect labels.
The incoming 11,847 audio clips before the refinement task featured a total of 13,681 labels, out of which 773 (5.7\%) were then rejected by the annotators. This means that 94.3\% of the incoming labels were verified as correct. This gives a sense of the level of label correctness obtained with the validation task (and therefore, the dev set).
In sum, after the validation task, for a pool of 11,847 audio clips mostly selected for eval (carrying a total of 13,681 labels), 94.3\% of the incoming labels were verified as correct, and 50.9\% of the incoming clips had one or more missing labels.
This indicates that in the dev set the labels are mostly correct, yet there is a large amount of clips featuring missing labels. However, the actual number of clips with missing labels in the dev set is estimated as less than 50.9\% because, as explained earlier, clips selected for dev tend to have less average number of sources per clip than those selected for eval.
% OLD
% We can use the refinement task processing to quantify the label noise at the output of the validation task.
% A total of 11,847 audio clips were processed with the refinement task, the majority of which ended up in the eval set.
% The processing undergone by these clips in this task in order to approach a complete sound event transcript is summarized next.
% A total of 6030 (50.9\%) received at least one additional label, indicating that there was some unlabeled material. For these 6030 clips, a total of 10,473 labels were generated. One must be careful when extrapolating these numbers to the dev set---as explained earlier, clips selected for dev tend to have, on average, less number of sources per clip.
% The incoming 11,847 audio clips featured a total of 13,681 labels, out of which 773 (5.7\%) were rejected by the annotators. This means that 94.3\% of the incoming labels were verified as correct. This gives a sense of the level of label correctness obtained with the validation task (and therefore, the dev set).
For the eval set, we have not quantified the amount of correctness and completeness due to lack of resources.
However, the amount of correctness is expected to be not very different from the 94.3\% estimated at the output of the validation task. 
We expect this because the hired raters are more qualified at the refinement task than during the previous validation task, as they have gained more annotation experience and a deeper knowledge of the ontology.
Labelling errors in FSD50K can be reported via its companion site.\textsuperscript{\ref{foot_companion}}
In this way, future dataset releases can include fixes reported in a collaborative way.

\subsubsection{Data Imbalance}
While some classes are abundant, others are much less represented due to the data scarcity in Freesound and/or low performance of the nomination system.
Another source of imbalance is the variable length of clips---some classes tend to contain shorter/longer clips depending on the sound events and the preferences of Freesound users when recording them.
Finally, the hierarchy of the ontology favours data imbalance between classes at different levels.

\subsubsection{Data Bias in Development Set}
% Because we prioritized the eval set over the dev set, 
Because we prioritized the allocation of small uploaders in the eval set to increase its diversity (Sec. \ref{ssec:split}), the development portion of a few classes is dominated by a few large uploaders.
Under the assumption that this signifies similar training examples in certain cases, this could create a data bias, which could be learned by models \cite{lei2011user}.  
This happens mainly in a few musical instruments, e.g., \textit{Trumpet}, due to the fact that Freesound users tend to upload many clips for these classes. 
Further analysis would be needed to determine if and how much this potential bias causes lack of generalization for these classes.

\subsubsection{Lack of Specificity in the Vocabulary}
Some leaf nodes in the ontology were merged to their parents due to data scarcity.
For instance, leaf nodes such as \textit{Blender}, \textit{Chopping (food)}, and \textit{Toothbrush} had to be merged with their parent \textit{Domestic sounds, home sounds}.
This motivated us to keep the latter class as a valid class despite that it is blocked in the ontology \cite{gemmeke2017audio}. 
% A natural extension of FSD50K is to grow these merged leaf nodes by adding more data.

\vspace{-2mm}
%============================================================================================
\subsection{Applications}
\label{ssec:apps}
FSD50K allows evaluation of approaches for a variety of sound recognition tasks. 
The most evident is multilabel sound event classification with large vocabulary \cite{Fonseca2019audio,fonseca2021shift}.
In this context, the proposed dataset supports several approaches such as learning sound event representations directly from waveforms \cite{cakir2016filterbank,park2020cnn}; analysis of label noise mitigation methods leveraging the non-exhaustive labeling of the dev set \cite{Fonseca2019learning,fonseca2020addressing,Fonseca2019model}; multimodal approaches using audio and text information (e.g., using the provided Freesound tags, title, and textual description for the clips) \cite{elizalde2019cross,favory2020coala}; evaluation of hierarchical classification via ontology-aware learning frameworks \cite{cramer2020chirping,shrivaslava2020mt,jati2019hierarchy}; or approaches specifically combining strong and weak labels \cite{kumar2017audio}.
By leveraging the common vocabulary between FSD50K and AudioSet, we hope that a number of tasks become possible, such as experimenting with domain adaptation techniques \cite{gharib2018unsupervised}, or cross-dataset evaluation \cite{bogdanov2016cross} under different acoustic conditions.
Other tasks include search result clustering in large vocabulary datasets \cite{favory2020search} or universal sound separation \cite{Kavalerov_WASPAA2019}.
%Finally, the proposed eval set can be used as a benchmark beyond FSD50K.
In addition, FSD50K has already accomplished several milestones.
A subset of the data has been used for a number of smaller datasets for classification \cite{fonseca2018general,Fonseca2019learning,Fonseca2019audio,abesser2021usm} and source separation \cite{Wisdom_InPrep2020}.
%, which add to previously existing datasets nourished from Freesound content \cite{salamon2014dataset,foster2015chime,piczak2015esc,stowell2013open,DCASE2017challenge}.
Likewise, subsets of FSD50K have enabled several sound recognition Challenges---specifically, DCASE 2018 Task 2 ``General-purpose tagging of Freesound audio with AudioSet labels" \cite{fonseca2018general}, DCASE 2019 Task 2 ``Audio tagging with noisy labels and minimal supervision" \cite{Fonseca2019audio} and Task 4 ``Sound event detection in domestic environments with  weakly labeled data and soundscape synthesis" \cite{turpault2019sound}, and DCASE 2020 Task 4 ``Improving  sound  event  detection in  domestic  environments  using  sound  separation" \cite{turpault2020improving}.
% These multiple contributions showcase the value of this effort.

\vspace{-2mm}
%============================================================================================
\subsection{FSD50K and AudioSet}
\label{ssec:compare}
As FSD50K and AudioSet are based on the same ontology and thus are partially compatible, we discuss the main similarities and differences between both. 
Table~\ref{tab:fsd_aset} summarizes some of them. 
\begin{table}[ht]
\vspace{-2mm}
\caption{Comparison of some properties of FSD50K and AudioSet}
\vspace{-2mm}
\centering
\begin{tabular}{lcc}
\toprule
             & \textbf{FSD50K} & \textbf{AudioSet} \\
\midrule
classes          & 200             & 527              \\
content          & waveform        & features        \\
%stability        & yes             & no             \\
dev clips        & 40,966          & $\approx$2M       \\
eval clips       & 10,231          & 20,383                  \\
%content license          & several CCs     & CC-BY-4.0        \\
clip length       & 0.3-30s        & $\approx$10s        \\
dev labeling      & CpI            & CpI             \\
eval labeling     & exhaustive     & CpI          \\
source           & Freesound audio  & Youtube video      \\
train/val split  & \checkmark      & -      \\
P.563 SNR (mean, median) [dB]      & (26, 25)  & (14, 10)      \\
%\midrule
\bottomrule
\end{tabular}
\label{tab:fsd_aset}
% \vspace{-1mm}
\end{table}
Both datasets use the AudioSet Ontology for organization, however FSD50K uses a smaller subset.
All classes in FSD50K are represented in AudioSet, except \textit{Crash cymbal} as well as four classes that are blocked in AudioSet but not in FSD50K (\textit{Human group actions}, \textit{Human voice}, \textit{Respiratory sounds}, and \textit{Domestic sounds, home sounds}).
%The reason to preserve these classes is that they aggregate clips from some of their children that were not considered large enough for the release.
The official AudioSet release consists of audio features pre-computed at 960ms-resolution, released under CC-BY-4.0 license.
FSD50K provides audio waveforms under several CC licenses as decided by Freesound users.
%Apart from this, some AudioSet users attempt to download the corresponding video soundtracks from YouTube.
In terms of stability, FSD50K is downloadable as several zip files from its Zenodo page.\textsuperscript{\ref{foot_zenodo}}
AudioSet features can be downloaded as a tar.gz file from the AudioSet website.\textsuperscript{\ref{foot_audioset_download}}
The original YouTube video soundtracks are gradually disappearing as they are subject to deletions and other issues, and their usage may be affected by copyright policies (Sec. \ref{sec:intro}).
% As seen in Table~\ref{tab:fsd_aset}, 
AudioSet's dev set is significantly larger than FSD50K's whereas AudioSet's eval set is roughly twice that of FSD50K.
Since AudioSet has a vocabulary 2.6 times larger, this means that in some classes there is more evaluation content in FSD50K.
Clips in AudioSet last $\approx$10s, whereas in FSD50K their length varies from 0.3 to 30s.
Thus, label weakness is more homogeneous in AudioSet, whereas it varies significantly in FSD50K, yielding quasi-strong labels as clips get shorter, and much weaker labels in the longest clips.

In terms of labeling, FSD50K provides event predominance annotations (``Present and predominant" \& ``Present but not predominant", Sec. \ref{ssec:task_val}) while AudioSet only provides presence annotations (``Present").
While it is not easy to objectively compare label quality in both datasets, we speculate that the labeling of both dev sets could be generally regarded as \textit{Correct but Potentially Incomplete} (CpI), i.e., both dev sets would be affected by a certain amount of missing labels. 
However, it seems reasonable to assume that, in the FSD50K portion of rather short sounds with PP annotations (see Sec. \ref{ssec:discussion}), the amount of missing labels is minimal.
This assumption is consistent with our observations from the annotation experience.
The eval set of FSD50K was exhaustively annotated; therefore, absence of labels means absence of sound events (except human error).
By contrast, the eval annotations in AudioSet would be in general CpI, similar to those of the AudioSet train set.
Unlike AudioSet, FSD50K consistently provides all relevant labels in a hierarchical path, except in a few specific cases of ambiguous ancestors.
As additional resources, we provide additional metadata (e.g., Freesound tags and class-wise annotation FAQs) and allow flagging labeling errors.\textsuperscript{\ref{foot_companion}}

Finally, despite both datasets being highly heterogeneous, we make the following conjectures.
Freesound clips are typically recorded with the goal of capturing audio, which is not necessarily the case in YouTube videos.
Additionally, given the AudioSet size, its audio clips are presumably recorded with a higher diversity of devices.
This would provide AudioSet with a higher diversity of audio qualities, often including more real-world and lower SNR conditions than Freesound audio (see Table \ref{tab:fsd_aset} for a rough SNR estimation described in Sec. \ref{ssec:discussion}).
Thus, a certain acoustic mismatch between both datasets may be expected.
In our view, both datasets suppose complementary resources for sound event research.

%% file: S5_Experiments.tex
In this Section, we conduct a set of multi-label sound event tagging (SET) experiments to give a sense of the performance that can be achieved with FSD50K using a baseline pipeline (Sec. \ref{ssec:baseline}), and to learn about the main challenges to consider when splitting Freesound audio for sound event recognition (SER) tasks (Sec. \ref{ssec:exp_val}).
For reproducibility, implementation details of evaluation metrics, learning pipeline, and networks can be inspected in the open-source code.\textsuperscript{\ref{foot_github}}

%============================================================================================
\subsection{Evaluation}
\label{ssec:metrics}

Some common evaluation metrics for SET (e.g., F-score or overall error ratio) depend on an operating point, i.e., a decision threshold applied on the per-class output scores.
These metrics encompass evaluation of the model's performance \textit{and} of the decision threshold tuning.
However, we believe that decoupling these two factors is desirable as, strictly, they are two different issues and the optimality of the latter can be application-dependent. 
Thus, we propose metrics able to evaluate a model's performance globally, integrating all possible operating points such that setting a decision threshold is not needed.
This trend has been adopted in other fields such as speaker recognition \cite{van2007introduction} and also recently in sound event detection (SED) \cite{bilen2020framework}.

%WC
On the one hand, we use common within-class metrics, i.e., metrics that rank all test samples according to the classifier score for one given class.
These metrics deal with only one classifier output at a time, such that calibration across different classifier outputs is irrelevant.
Following \cite{gemmeke2017audio,hershey2017cnn}, we use mean Average Precision (mAP) and $d'$.
mAP is the mean across classes of the Average Precision (AP), which summarises the precision-recall (PR) curve as the classifier decision threshold is varied.
%obvio lo q es PR curve. AP summarises the precision-recall (PR) curve, which is defined as the relation between the precision and the recall as the classifier discrimination threshold is varied \cite{scikit_PR}.
% weird and non intuitive: AP is calculated as the weighted average of precisions attained at each threshold, with the weight being the increase in recall from the previous threshold, such that thresholds considered are those at which a new positive sample is recalled .
%from Hershey: which is the proportion of positive items in a ranked list of trials (i.e., Precision)averaged across lists just long enough to include each individual pos-itive trial [28]. 
AP is calculated as the Precision (i.e., the proportion of positive samples in a ranked list) averaged across all the lists just long enough to recall a new positive sample \cite{scikit_PR,hershey2017cnn}.
AP is very similar to the area under precision-recall curve (PR-AUC), both being the most common ways of summarising a PR curve---the difference between them lies in implementation details \cite{scikit_metrics,VLFeat}. 
$d'$ (d-prime) measures the separation between the means of two unit-variance normal distributions (corresponding to the scores for positive and negative examples) that would achieve the same ROC-AUC \cite{green1966signal}.
$d'$ is computed as a monotonic transform of ROC-AUC \cite{green1966signal,hershey2017cnn}:
\begin{equation}
\label{eqn:dprime}
d'=\sqrt{2} F^{-1}(ROCAUC),
\end{equation}
% EQUATION: d’ = sqrt(2) Z(ROC AUC)
% d′ = √2F −1(AUC) 
where $F^{-1}$ is the inverse cumulative distribution function for a unit Gaussian.
% More details about $d'$ can be found in \cite{green1966signal}.
%BC OLD simple version 
% To complement the within-class metrics, we propose to use a between-class metric, i.e., which evaluates the overall ranking across all classifier outputs for every test sample.
% Specifically, we use \textit{Label-weighted label-ranking average precision} (abbreviated as {\lwlrap} and pronounced ``lol wrap''), which was recently introduced for DCASE 2019 Task 2 \cite{Fonseca2019audio}.
% {\lwlrap} measures, for every ground truth test label $c$, what fraction of the predicted top-ranked labels down to $c$ are among the ground truth.
%More details about \lwlrap can be found in \cite{Fonseca2019audio}.
%BC NEW version with equation, v¡for reviewer 
To complement the within-class metrics, we propose to use a between-class metric, i.e., which evaluates the overall ranking across all classifier outputs for every test sample.
Specifically, we use \textit{Label-weighted label-ranking average precision} ({\lwlrap}), recently introduced for DCASE2019 Task 2 \cite{Fonseca2019audio}.
% {\lwlrap} measures, for every ground truth test label $c$, what fraction of the predicted top-ranked labels down to $c$ are among the ground truth.
Let $C(s)$ be the set of reference labels for test sample~$s$, $Prec(s, c)$ be the label-ranking precision for the list of labels up to class $c$, and $Rank(s, c)$ be the rank of class label $c$ in that list.
$Prec(s, c)$ is equal to 1 if all the top-ranked labels down to $c$ are part of $C(s)$, and at worst case equals $1/Rank(s, c)$ if none of the higher-ranked labels are correct. Then, {\lwlrap} can be expressed by
\begin{equation}
\label{eqn:lwlrap}
    l\omega lrap = \frac{1}{\sum_s |C(s)|} \sum_s \sum_{c \in C(s)} Prec(s, c)
\end{equation}
where $|C(s)|$ is the number of reference labels for sample~$s$.
A Python implementation of {\lwlrap} is provided online.\footnote{\url{https://colab.research.google.com/drive/1AgPdhSp7ttY18O3fEoHOQKlt_3HJDLi8}}
%final
For all metrics, larger is better.
mAP $\in [0, 1]$, non-pathological $d’ \in [0, \infty)$, and {\lwlrap}  $\in [0, 1]$.
%n.b.: negative d-prime is possible for worse-than-random classifiers.
All metrics are computed on a per-class basis, then averaged with equal weight across all classes to yield the overall performance (i.e., \textit{balanced} a.k.a. \textit{macro} averaging), as in \cite{hershey2017cnn,fonseca2020addressing,gemmeke2017audio}.

%============================================================================================
\subsection{Baseline Systems}
\label{ssec:baseline}
\begin{table*}[ht]
\vspace{-4mm}
\caption{Evaluation performance for the architectures considered}
\vspace{-2mm}
\centering
\begin{tabular}{l|cccccc}
\toprule
\textbf{Model} & \textbf{lr} & \textbf{weights}   & \textbf{mAP}  & $\bm{d'}$  & \textbf{\lwlrap}  \\
\midrule
CRNN      &  5e-4        & 0.96M       & 0.417 $\rpm$ 0.003   & 2.068 $\rpm$ 0.015   & \textbf{0.519}  $\rpm$ 0.002\\
VGG-like  &  3e-4        & 0.27M       & \textbf{0.434} $\rpm$ 0.002   & \textbf{2.167} $\rpm$ 0.011   & 0.514 $\rpm$ 0.003    \\
ResNet-18  &  1e-5       & 11.3M       & 0.373 $\rpm$ 0.001   & 1.883 $\rpm$ 0.020   & 0.465 $\rpm$ 0.001 \\
DenseNet-121  &  5e-5    & 12.5M       & 0.425 $\rpm$ 0.002   & 2.112 $\rpm$ 0.032   & 0.505 $\rpm$ 0.004    \\
%\midrule
\bottomrule
\end{tabular}
\label{tab:results}
\vspace{-4mm}
\end{table*}
%====================
Next, we benchmark several commonly used deep networks on the proposed FSD50K.

\subsubsection{Learning Pipeline}
Incoming audio is downsampled to 22.050 kHz and transformed to 96-band, log-mel spectrogram as input representation.
To deal with the variable-length clips, we use time-frequency (T-F) patches of 1s (equivalent to 101 frames of 30ms with 10ms hop)---thus the input to all models is of shape TxF=101x96.
Clips shorter than 1s are concatenated until such length, while longer clips are sliced in T-F patches with 50\% overlap inheriting the clip-level label (a.k.a. false strong labeling \cite{morfi2018data}).
%learning strategy
We adopt the train/val split designed in App. \ref{app:post_pro}.
We implement a learning pipeline in TensorFlow \cite{tensorflow2015-whitepaper}.
Models are trained using Adam optimizer \cite{kingma2014adam} to minimize binary cross-entropy loss, with initial learning rate depending on the network (see Table \ref{tab:results}), which is halved whenever the validation PR-AUC plateaus for 5 epochs (no tolerance).
Models are trained up to 100 epochs, earlystopping the training whenever the validation PR-AUC is not improved in 10 epochs.
We use a batch size of 64 and shuffle training examples between epochs.
Once the training is over, the model checkpoint with best validation PR-AUC is selected to predict scores and evaluate performance on the eval set.
We optimize PR-AUC (instead of other metrics based on ROC curves) because PR curves can be more informative of performance when dealing with imbalanced datasets \cite{davis2006relationship}.
Likewise, we use PR-AUC (instead of mAP) for simplicity as it is a built-in metric in TensorFlow.
%inference
For inference, we pass each (eval or val) T-F patch through the model to compute output scores, which are then averaged per-class across all patches in a clip to obtain clip-level predictions, as in \cite{fonseca2020addressing,gemmeke2017audio}.
We note this aggregation must be done also for validation---preliminary experiments validating at patch-level using inherited clip-level labels revealed misleading results.
Extensive hyper-parameter tuning (beyond learning rate) is not conducted.

%====================================================
\subsubsection{Network Architectures}

Current trends in SER encompass mainly CNNs \cite{hershey2017cnn,fonseca2021shift,kong2019panns,fonseca2017acoustic,fonseca2018simple} and CRNNs \cite{cakir2017convolutional,perez2019hybrid}.
We run experiments with the following networks, all of them ending with a fully connected layer of 200 units (the vocabulary size) with sigmoid activation to support multi-label classification.
Main parameters for CRNN and VGG-like architectures are set via non-exhaustive preliminary experiments.

\textbf{CRNN.}
This is one of the most used architectures for SED \cite{cakir2017convolutional}, and to a lesser extent for SET \cite{Ebbers2019}.
Our model, inspired by \cite{cakir2017convolutional}, has three convolutional layers of 128 filters with a receptive field of (5,5), each of them followed by Batch Normalization (BN) \cite{ioffe2015batch}, ReLU activation and max-pooling.
The max-pooling sizes are $(t,f) =$ (2, 5), (2, 4) and (2, 2)---since we are not interested in detecting events' timestamps, we pool also in the time dimension which reduces dimensionality without harming performance in our experiments.
To model events' temporal structure in the incoming feature maps, the convolutional stack is followed by a bidirectional GRU layer of 64 units, returning the last output of the output sequence.

\textbf{VGG-like.}
VGG-based architectures have been widely used for both SET \cite{fonseca2021shift,dorfer2018training} and SED \cite{kim2019sound}.
We use a model inspired by the original architecture \cite{simonyan2014very} from computer vision, but reduced to a much smaller size.
In particular, this model has three convolutional layers of 32 filters, two convolutional layers of 64 filters, and one convolutional layer of 128 filters.
All convolutional layers have a receptive field of (3,3) and are followed by BN and ReLU activation.
Between each group of convolutional layers with same number of filters, max-pooling of size (2,2) is applied.
Output feature maps are summarized by concatenating global max pooling and global average pooling per channel.
Summarizing the learnt audio representation via combination of these two poolings provided a small mAP boost with respect to using either of them individually.
Then, the outcome is passed through a fully connected layer of 256 units.

% CV standard
Finally, we also experiment with two architectures taken off-the-shelf from the computer vision literature. 
While the two previous networks received some tuning in their design, the next ones are the original architectures without any tuning whatsoever---only the input/output shapes to match our task.
\newpage
\textbf{ResNet-18.}
ResNets \cite{he2016deep} have been sucessfully used for SER \cite{jansen2018unsupervised,kong2019panns,fonseca2020addressing}.

\textbf{DenseNet-121.}
DenseNets are reported to outperform ResNets for image recognition \cite{huang2017densely}, and have been recently used for SET \cite{Fonseca2019model,iqbal2020learning}.

%====================
\subsubsection{Results}
Table \ref{tab:results} lists the results for the considered architectures, along with the learning rates used (after tuning) and the number of weights. 
% Each experiment trial is run three times with different seeds.
% We report average evaluation performance and standard deviation across trials.
Each network is trained from scratch three times with different random initialisation and different orderings in the training data.
We report average and standard deviation of the evaluation performance across the three trials.
The following results and discussion are based on the particular train/val/eval split used.
Interestingly, the best overall model across all metrics is VGG-like, despite being less modern and more lightweight than the other architectures.
This result accords with similar recent findings in music genre recognition \cite{won2020evaluation}.
The building blocks of VGG-like are very similar to those of the \textit{CNN14} network in \cite{kong2019panns}, which rivals state-of-the-art results in AudioSet classification \cite{kong2019panns}. 
However, CNN14 is much deeper and heavier (81M weights). 
The VGG-like model is closely followed by DenseNet-121, which has many more weights, and then by the CRNN, which shows the best {\lwlrap}.
CRNN architectures are also used in some top SED systems, e.g., in recent DCASE Challenge Task 4 editions \cite{turpault2019sound,turpault2020improving}.
ResNet-18 is found to be the worst performing model.
Curiously, we also observe that the optimal learning rates tend to be rather low for this architecture.
We also tried ResNet-34 in preliminary experiments, obtaining similar results (at the expense of many more weights).
Our results contrast with the successful results of \cite{kong2019panns,ford2019deep}, which achieved state-of-the-art performance for AudioSet classification using ResNet architectures.
Factors possibly influencing this different behaviour include the different amount of training data (much larger in AudioSet) or the data itself.
Results in Table~\ref{tab:results} suggest that, for our scale of data, smaller models with basic tuning and audio-informed design choices can outperform much larger off-the-shelf computer vision architectures; however, DenseNet-121 with no tuning provides good performance.

\begin{figure*}[h!]
%\begin{figure}[ht]
    %\vspace{-4mm}
    \centering
  \centerline{\includegraphics[width=1.15\textwidth]{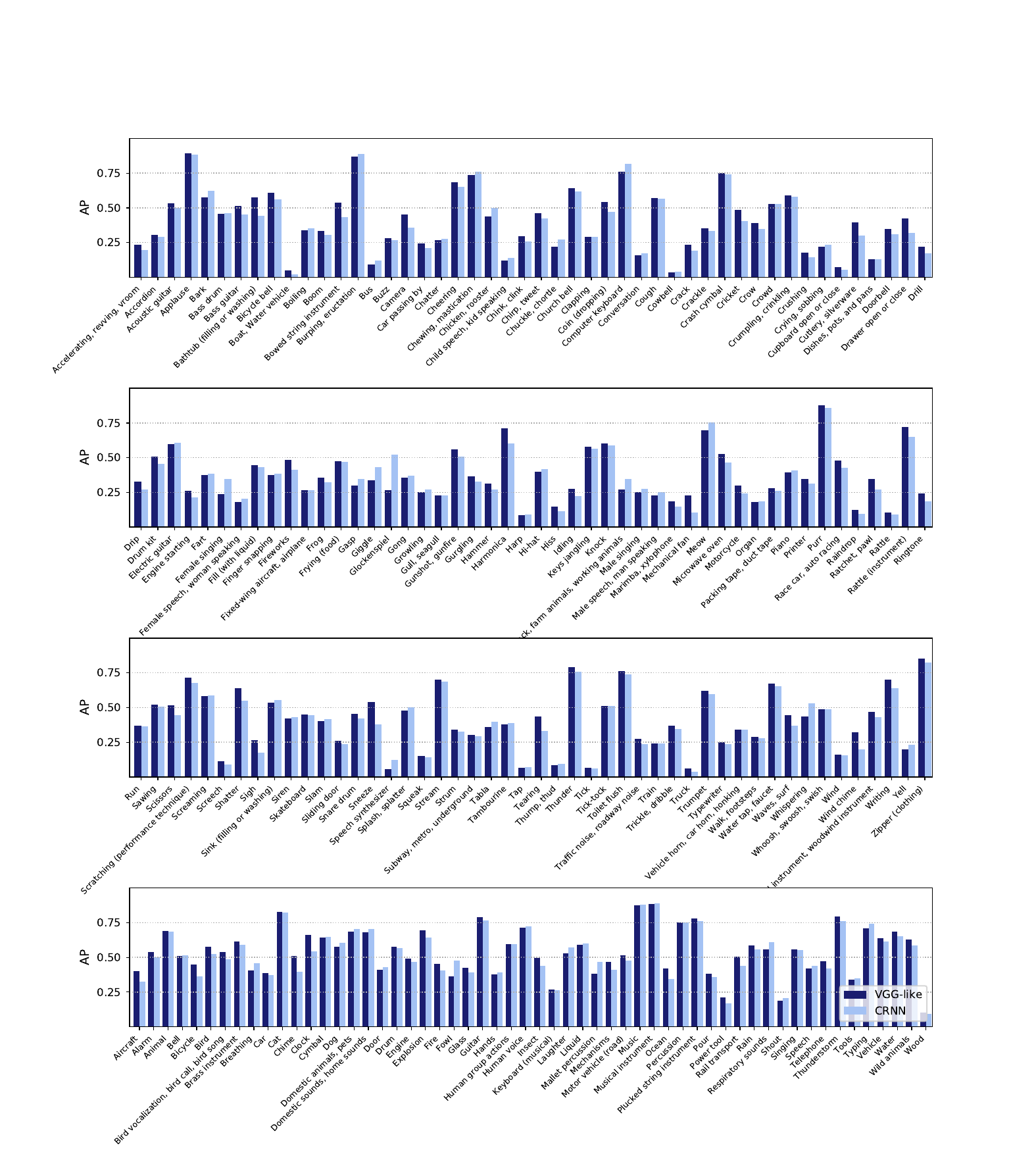}}
    \caption{Per-class average precision for all classes in FSD50K, using the best-performing VGG-like model (dark blue) and the CRNN model (light blue). Top 3 rows show the 144 leaf nodes and bottom row comprises the 56 intermediate nodes.}
    \label{fig:perclass}
    %\vspace{-4mm}
\end{figure*}

Fig. \ref{fig:perclass} shows the per-class AP (averaged across three trials) for all classes in FSD50K, using the best-performing VGG-like model (dark blue), and the CRNN model (light blue).
Leaf nodes with top recognition include \textit{Applause}, \textit{Burping, eructation}, \textit{Purr}, and \textit{Computer keyboard}, with AP over 0.75.
The worst performance is shown in \textit{Boat, Water vehicle}, \textit{Cowbell}, \textit{Speech synthesizer}, \textit{Tap} and \textit{Tick}.
After inspection of the latter classes, we conjecture this is due to aspects such as high intra-class variation, confound with other similar classes, ambiguity in the class definitions, or very short length of sound events---all of them being relevant challenges in SER.
% CRNN discussion
Finally, it can be seen that most per-class APs by the CRNN are slightly lower than those of the VGG-like model---as expected since VGG-like has a higher overall mAP.
However, there are a few exceptions in which the CRNN performs better, such as in different types of speech (either spoken, sung, screamed, yelled or whispered).
This is interesting as CRNNs were originally proposed for speech recognition \cite{sainath2015convolutional} before being adapted for SER \cite{cakir2017convolutional}.
Other exceptions include some human sounds (e.g., types of laughter (\textit{Chuckle, chortle} or \textit{Giggle}), \textit{Gasp}, or \textit{Crying, sobbing}), as well as some animal vocalizations (e.g., \textit{Bark}, \textit{Meow} or \textit{Chicken, rooster}).
These sounds share a marked temporal behaviour.
These results highlight the different behaviour, for some classes, of a model including a recurrent layer with respect to another relying only on convolutional layers.

\vspace{-3mm}
%============================================================================================
\subsection{Impact of Train/Validation Separation}
\label{ssec:exp_val}
In App. \ref{app:post_pro} we discuss some factors to consider when splitting Freesound audio data for machine learning, and we design a validation set emphasizing the issue of data contamination.
Here, we experimentally analyze the impact of contamination in this setting.
To this end, we pick one architecture from Sec.~\ref{ssec:baseline} (the CRNN) and we train and evaluate it utilizing three different train/validation splits in order to show their differences.
Specifically, let us consider three candidate validation sets obtained with different approaches:
\begin{enumerate}
    \item \textbf{val\_random} is computed via \textit{random} sampling. We run 3000 trials of a train/validation separation and we select the validation set with minimum Jensen-Shannon (JS) divergence\footnote{The JS divergence is based on the Kullback-Leibler divergence but it is symmetric. We use it as a distance metric to measure similarity between the development and validation distributions, similarly as in \cite{cramer2020chirping}.} with respect to the development set.
    \item \textbf{val\_is} is computed via \textit{iterative stratification} \cite{sechidis2011stratification}. We run 3000 trials of a train/validation separation and we select the validation set with the minimum number of shared uploaders between training and validation.\footnote{Minimizing the JS divergence is not needed here as stratification is already the objective of this method, hence all separations have a fairly consistent JS divergence.}
    \item \textbf{val} is the validation set proposed in App.~\ref{app:post_pro}.
\end{enumerate}

% In all cases, the validation set is initialized with most of the data transferred from evaluation to development (App.~\ref{app:post_pro})---since this content is exhaustively labeled, it is well suited for evaluation purposes.
In all cases, the validation set is initialized with most of the data that was transferred from the first evaluation set prototype to the development set for balancing purposes (App.~\ref{app:post_pro}). 
Since this content is exhaustively labeled, it is well suited for evaluation purposes.
The main characteristics of the three validation sets are listed in Table~\ref{tab:exp_val}.
\begin{table}[ht]
\vspace{-1mm}
\caption{Main statistics for the considered validation sets}
\vspace{-2mm}
\centering
\begin{tabular}{@{}l|ccc@{}cc@{}}
\toprule
\textbf{Validation Set}        & \textbf{clips} & \textbf{duration}  & \textbf{JS}  & \textbf{shared}    & \textbf{PR-AUC}  \\
                   &                  &                  &             & \textbf{uploaders}  & \textbf{drop}               \\
\midrule
\textit{val\_random}      &  4697              & 9.7h      & $1.8\times10^{-2}$  & 930             & 0.15    \\
\textit{val\_is}        &  4543              & 9.3h        & $6.8\times10^{-3}$  & 857             & 0.14    \\
\textit{val} (proposed)  &  4170              & 9.9h       & $2.1\times10^{-2}$  & 641             & $\approx0$    \\
%\midrule
\bottomrule
\end{tabular}
\label{tab:exp_val}
\vspace{-4mm}
\end{table}
The sets \textit{val\_random} and \textit{val\_is} amount to $\approx$15\% of the development data associated with leaf nodes (Table \ref{tab:main_stats}); \textit{val} is slightly lower (13.3\%) due to allocating less validation data for the most abundant classes as well as some approximations (App.~\ref{app:post_pro}).
All validation sets have a similar duration. 
In terms of stratification, the split done through iterative stratification, \textit{val\_is}, yields more similar class distributions than the other two, which are on par.
The main differences lie in the uploaders “shared” between train and validation, both in number and in their nature.
In particular, \textit{val\_random} and \textit{val\_is} suffer from within-class (WC) and between-class (BC) contamination as no measure is taken to prevent them.
By contrast, \textit{val} is designed to minimize WC contamination while being relatively flexible with BC contamination (see App.~\ref{app:post_pro} for definitions of WC and BC contamination, as well as for the design of \textit{val}).
Therefore, not only is the number of shared uploaders less in the proposed \textit{val}, but also the contamination is limited mostly to BC.
\begin{figure*}[ht!]
%\begin{figure}[ht]
    \vspace{-2mm}
    \centering
  \centerline{\includegraphics[width=1.00\textwidth]{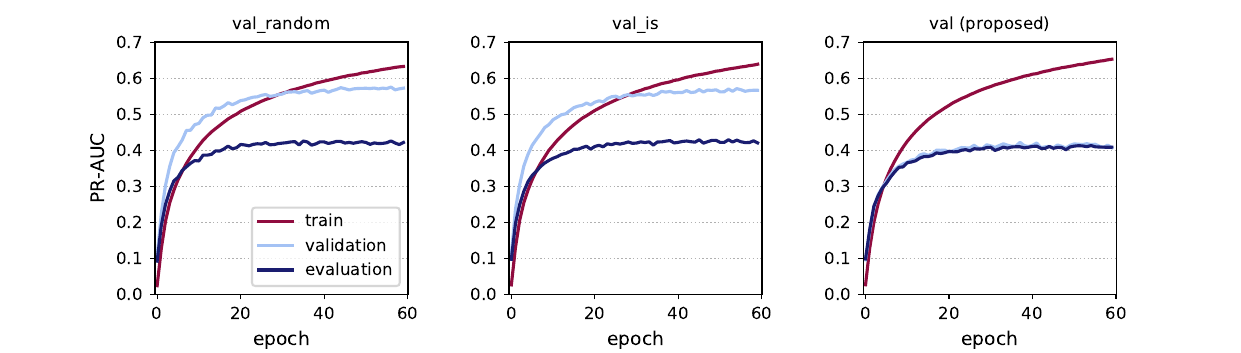}}
    \vspace{-1mm}
    \caption{Learning curves (PR-AUC for train, validation, and evaluation) for the CRNN model using the three train/validation splits specified in Table~\ref{tab:exp_val} (\textit{val\_random} (left), \textit{val\_is} (middle), and the proposed \textit{val} (right)). Validation performance is substantially better than evaluation performance when using \textit{val\_random} and \textit{val\_is}, in which the classifier is trained and validated on clips from the same uploader \textit{and} the same class (WC contamination). When this type of contamination is minimized (\textit{val}), validation performance is a good proxy of evaluation performance.}
    \label{fig:exp_val}
    \vspace{-2mm}
\end{figure*}

To compare the candidate splits, we train the CRNN of the previous Section using the three of them (in this case, with a learning rate of 1e-4 and no learning rate scheduling).
Fig.~\ref{fig:exp_val} illustrates the learning curves (PR-AUC for train, validation, and evaluation) using each of the splits.
We display 60 training epochs allowing validation and evaluation performance to roughly stabilise.
From Fig.~\ref{fig:exp_val} and Table~\ref{tab:exp_val} several observations can be made.
In the left and middle plots of Fig.~\ref{fig:exp_val} (corresponding to \textit{val\_random} and \textit{val\_is}), validation performance is substantially better than evaluation performance.
In these cases, the classifier is trained and validated on clips from the same uploader \textit{and} the same class (i.e., WC contamination).
We call this the ``uploader effect" (following the analogy of the ``album effect" \cite{mandel2005song} or ``artist effect" \cite{flexer2009album}).
% The left and middle plots of Figure~\ref{fig:exp_val} show the “uploader effect” (following the analogy of the “album effect” \cite{mandel2005song} or ``artist effect" \cite{flexer2009album}) in which the classifier performs significantly better on validation when trained and validated on clips from the same uploader \textit{and} the same class (WC contamination).
In Table~\ref{tab:exp_val}, it can be seen that the number of uploaders shared between train and validation is positively correlated with the validation-evaluation PR-AUC drop.
In the cases of \textit{val\_random} and \textit{val\_is} we observe substantial performance drops, whereas with \textit{val} the performance drop is very small.
%1
Results from Fig.~\ref{fig:exp_val} and Table~\ref{tab:exp_val} suggest that, when contamination is considered and minimized, validation performance is a good proxy of evaluation performance---otherwise, it can be overly optimistic.
Consequently, if the model is tuned using the validation set it may occur that, depending on the type and amount of contamination, the tuning reflects model's ability to partially overfit train data rather than to generalise to unseen data.
%2
In addition, our results indicate that the distinction between WC and BC contamination seems reasonable in the context of Freesound audio organized with a large vocabulary, confirming our initial hypothesis that WC is the most harmful type while BC has lesser impact (see App.~\ref{app:post_pro}).

%3
Lastly, we observe a slightly higher train performance and slightly lower validation and eval performances when using \textit{val} (right plot of Fig.~\ref{fig:exp_val}), which content comes mostly from a variety of small uploders.
Under the assumption that not all training examples are equally informative (which is the basis for disciplines like instance selection \cite{liu2002issues}), this may occur because the content transferred to \textit{val} includes some highly informative examples.
% Yet, we propose this train/validation split as we deem it more methodologically correct than the others for systems’ benchmarking.
Yet, we propose this train/validation split for systems’ benchmarking because we deem it more methodologically correct than the others given that data contamination is minimized. 
In summary, carefully splitting Freesound audio is important as it can have a non-negligible impact on learning and performance. 
Therefore, for reproducibility and fair comparability of results, system benchmarking should be done explicitly specifying the validation split that was used.

%ADD: The insight confirmed by these results is relevant as, to our knowledge, none of the previous works presenting datasets nourished from Freesound audio address explicitly the issue of user contamination.
%based on 
%composed of

%% file: S6_Conclusion.tex
%We have included an in depth description of the FSD50K creation process, as well as a comprehensive dataset characterization and baseline SET experiments.

%summary
In this paper, we introduced \textit{FSD50K}, a dataset containing 51,197 Freesound clips totalling over 100h of audio manually labeled using 200 classes drawn from the AudioSet Ontology.
The audio clips are CC-licensed, thereby making the dataset freely distributable (including audio waveforms).
%conclusions
We proposed a methodology for creating datasets of sound events based on human validation and refinement, and using a mixture of crowd-sourcing and recruited trained annotators.
In this process, we experienced how human labeling of a large vocabulary of everyday sounds is a laborious and complex task.
Special emphasis was put on the careful curation of the evaluation set content and labels, so that it can serve as a reliable evaluation benchmark.
We showed how it is important to adapt the dataset creation process to the specifics of the source data---in our case, Freesound audio and metadata, and the AudioSet Ontology---and how a deep knowledge of these data is crucial to identify data challenges and limitations, and to avoid pitfalls in the creation of the dataset.
Finally, through experimental results we showed that, for FSD50K classification, smaller models with basic tuning and audio-informed design choices can outperform larger off-the-shelf computer vision architectures.
We also showed that within-class data contamination must be considered when splitting Freesound audio as it can have a considerable effect on the evaluation of sound event classifiers.
%final words
%complementary to AudioSet
%old. FSD50K aims to be an open and stable alternative to complement AudioSet to stimulate reproducible SER research.
%FSD50K is an open and stable dataset that aims to complement AudioSet to foster reproducible large-vocabulary SER research.
FSD50K is an open and stable dataset aimed at complementing AudioSet in order to foster reproducible large-vocabulary SER research.
%FW
In the future, dataset extensions could be carried out.
More data could be added via semi-automatic methods by leveraging models trained on FSD50K to scale up efficiently.
Likewise, the vocabulary could be extended by growing the merged leaf nodes in FSD50K.

%% file: S9_Appendices.tex
\vspace{-3mm}
%========================================================================================
\section{Ontological Nomenclature}
\label{app:onto}
%Similarly, FSD50K is also organized using a subset of the AudioSet Ontology.\footnote{In the following, we shall sometimes refer to the AudioSet Ontology as the ontology.}
We clarify next some basic (albeit relevant) ontology-related terms used in this paper.
%The 632 classes in the ontology are commonly regarded as \textit{taxa} in the taxonomic nomenclature. 
We shall refer to the 632 classes in the ontology as \textit{nodes} (either \textit{leaf} nodes when they are located at the very bottom of the hierarchy, or \textit{intermediate} nodes otherwise).
We shall also use the ontological terms \textit{children} and \textit{parents}, as widely used in ontology-related genome research \cite{jantzen2011go}.
Note that, by definition, leaf nodes do not have children nodes, while the intermediate nodes do.
Similarly, given a node, we refer to all the parent nodes connecting it to the root of the ontology as \textit{ancestors}.
As an example, let us consider the hierarchical path: \textit{Root} $\rightarrow$ \textit{Natural sounds}  $\rightarrow$ \textit{Thunderstorm} $\rightarrow$ \textit{Thunder}. 
In this path, \textit{Thunder} is a leaf node; \textit{Natural sounds} and \textit{Thunderstorm} are both intermediate nodes; \textit{Thunderstorm} is child of \textit{Natural sounds} and parent of \textit{Thunder}; and \textit{Thunderstorm} and \textit{Natural sounds} are all the ancestors of \textit{Thunder}.

\vspace{-2mm}
%========================================================================================
\section{Post-processing Stage for FSD50K}
\label{app:post_pro}
The outcome of the refinement task (Sec. \ref{ssec:task_gen}) is two sets of data: a candidate development set with correct but potentially incomplete labels, and an exhaustively-labeled candidate evaluation set, using a vocabulary of 395 classes.
Here, we describe the post-processing carried out to obtain what's finally released as FSD50K, using a vocabulary of 200 classes.
% This process involves, mainly, merging low prior classes into their parents thus ensuring a minimum amount of per-class data. 
% This is what we define as \textit{ground truth} for FSD50K, with a vocabulary of 200 classes. 

\vspace{-3mm}
\subsection{Determine FSD50K vocabulary}
%We define valid leaf nodes as those with at least 100 clips, and without an extreme development/evaluation imbalance. 
We define \textit{valid} leaf nodes as those meeting two requirements: a minimum of 100 clips and without extreme development/evaluation imbalance. 
This is a trade-off between abundant per-class data and preserving a lot of leaf nodes.\footnote{Given the particularities of some classes, the requirements to consider a leaf node \textit{valid} are relaxed in a few exceptions.} We take the following measures.

\textbf{Merge non-valid leaf nodes with their parents.}
%First, non-valid leaf nodes are merged with their parents. 
There are two variants of this process, depending on the type of branch in the hierarchy.
First, non-valid siblings of valid leaf nodes are merged with their parents.
In these branches, the level of specificity is fixed by the valid sibling.
For instance, \textit{Yip}, a class with few data which is sibling of \textit{Bark} and child of \textit{Dog}, is merged with \textit{Dog} and the most specific label in this branch is the valid leaf \textit{Bark}.
Then, in branches without any valid leaf nodes, all leaf nodes are merged with their parents, which in turn become new leaf nodes (since they no longer have children).
This process is repeated recursively, pruning the branch by moving upwards in the hierarchy, until a new leaf node becomes valid.
While we ideally want to prune the branches as little as possible to preserve the most specific nodes, some low-level nodes are inevitably merged with non-specific parents, e.g., \textit{Domestic sounds, home sounds}.
The minimum data requirement is enforced at the leaf node of every branch, but not at its ancestors, which are intrinsically valid because the leaf node provides enough data. This means that, occasionally, the data explicitly associated with one ancestor may be scarce. 
This is due to the nomination system and annotation processes, which favour more specific labels.
%For example, \textit{Rail transport} is a valid intermediate node despite it holding very few data, because their children (\textit{Train} and \textit{Subway, metro, underground}) are valid, holding most of the data in this branch. 

\newpage
\textbf{Remove some valid leaf nodes to obtain a more semantically consistent vocabulary.}
%Second, we remove some valid leaf nodes to obtain a more semantically consistent vocabulary.
As a result of the pruning, some parents with various children in the ontology end up having very few children in the candidate dataset.
In most cases, this is not a problem as children are rather independent semantically.
%---e.g., \textit{Domestic sounds, home sounds} encompasses 27 children that occur usually in domestic contexts but with weak semantic links otherwise.
However, in other cases, children constitute a pre-established subset of closely related classes that makes more sense when all of them co-exist, e.g., the classes \textit{Light engine (high frequency)}, \textit{Medium engine (mid frequency)}, and \textit{Heavy engine (low frequency)}, where only the former is valid.
Considering the real operation of trained models, the fact that only one of these children is valid could potentially lead to unnatural predictions biased by the choice of the vocabulary.
%, i.e., recurrently spiking \textit{Light engine (high frequency)}, absent their complementary siblings.
To prevent this issue, we merge some ``isolated" valid leaf nodes with their parents.
%, thus obtaining a more semantically consistent vocabulary. 
%Apart from the mentioned engine types, this also occurs with wind and bowed string instruments.
Hence, despite having a substantial number of light engine sounds, they are not part of the vocabulary---only \textit{Engine} is. Note however that these more specific annotations are indeed available in the sound collection format.

\textbf{Discard some intermediate nodes.}
%Lastly, we discard some non-essential intermediate nodes:
This includes classes of abstract nature or with ambiguous children and few data, e.g., \textit{Digestive} or \textit{Arrow}, respectively. The outcome is a vocabulary of 200 classes (144 leaf nodes and 56 intermediate nodes).

\vspace{-2mm}
\subsection{Balancing development/evaluation sets}
%=========================================================================
As a result of exhaustively labelling the evaluation set, the proportion of some frequently occurring sound events increased substantially, sometimes exceeding the number of labels in the development set.
To obtain a better balance between development and evaluation sets, we first identified a set of 40 leaf nodes which benefit from transferring data from evaluation to development.
Then, we selected a set of evaluation clips such that: \textit{i)} their content encompasses mainly the 40 target classes with a minimal impact on the remaining ones---note the clips are multilabel; \textit{ii)} they are disjoint from the remaining set of clips in terms of uploaders. %as explained in Sec. \ref{ssec:split}.
%Experiments were carried out to confirm that this data removal did not cause any pathological issue on the sound classes.
Specifically, we transferred 1182 clips, resulting in an evaluation set of 10,231 clips, and a per-class development/evaluation proportion ranging from 50/50\% to 75/25\% in the vast majority of leaf nodes.
The per-class split proportion depends on data availability, ubiquity of the sound events, degree of multilabelness of the audio clips, and non-divisibility of content from the same uploader.
Exceptions include \textit{Chatter}, \textit{Chirp, tweet} and \textit{Male speech, man speaking}, for which there are more evaluation than development labels due to the exhaustive labelling of these ubiquitous events.
With this transfer, we also make available some exhaustively labeled content for validation.

\vspace{-3mm}
%=========================================================================
\subsection{Validation Set}
Some recent large audio datasets do not provide predefined validation sets \cite{gemmeke2017audio,Mesaros2018_DCASE} allowing dataset users to create their own.
Nonetheless, for easier dataset consumption and reproducibility we propose a candidate split of the development set into train and validation.
We consider that a validation set should ideally meet the following criteria:
\begin{itemize}
    \item \textbf{Proportion}. The validation set typically amounts to a given proportion of the development set, often between 10 and 20\%. Note that due to the multilabel and variable-length nature of Freesound audio, the proportion can be different in terms of audio clips, labels, and duration.
    \item \textbf{Stratification}. It is usually desirable that the class label distribution is similar in both train and validation sets.
    %, that is, the frequency of occurrence of each label is roughly preserved between both sets.
    \item \textbf{Contamination}. As explained in Sec. \ref{ssec:split}, contamination across splits should be minimized.
\end{itemize}
%Unfortunately, due to several practical constraints, these criteria can not always be met, and our case is not an exception.

Typical ways to make train/validation splits include random sampling or iterative stratification \cite{sechidis2011stratification}.
Both can produce desired data proportions and class distributions, the latter being popular for multilabel data.\footnote{Random sampling does not account for stratification per se, but a workaround is to compute many train/validation splits and choose the one that minimizes a distance between the respective class distributions.} 
However, they fail to keep non-divisibility of uploaders' content, thus generating contamination.
The distribution of number of clips per uploader is very varied in the development set.
However, since we already allocated a large amount of small uploaders into the evaluation set (Sec. \ref{ssec:split}), preserving uploader non-divisibility at this point means deviating from the target class distribution.
In other words, it is difficult to strictly meet the three above criteria simultaneously, hence we need to relax their application.

We focus on the contamination criteria and distinguish two types of contamination: \textit{i)} \textit{within-class} contamination (WC, when content from the same uploader \textit{and} belonging to the same class is placed at both train and validation sets);
\textit{ii)} \textit{between-class} contamination (BC, when content from the same uploader \textit{but not} from the same class is placed at both train and validation sets).
We hypothesize WC is more harmful as it could imply having the same sound source, physical location and/or recording gear in both sets. 
By contrast, BC would have less impact as, in most cases, the audio material would be different, and also possibly the acoustic environment.
Under this hypothesis, we focus on minimizing WC contamination while being flexible with BC.
To do this, we employ a method similar to that of Sec. \ref{ssec:split}.
We first define the content from one uploader labeled with the same class label as the minimum non-divisible unit.
Then, we adopt an iterative process in which, after sorting the uploaders per-class appropriately, we progressively allocate their content to the validation set.

As preprocessing, we initialize the validation set with most of the data transferred from evaluation to development---this content is well suited for evaluation purposes as it is exhaustively labeled. 
We then compute a score per uploader and per class. 
The score for uploader $u$ in class $c_i$ is given by:
%\ref{eqn:score_val}
% \vspace{-1mm}
\begin{equation}
\textrm{score}_{c_{i}}^u = \alpha \textrm{n\_labels}_{c_{i}}^u + \beta \frac{1}{K_u} \sum_{k=1}^{K_u} \textrm{n\_labels}_{c_{k}}^u,
\label{eqn:score_val}
\end{equation}
where $\textrm{n\_labels}_{c_{i}}^u$ represents the number of labels provided by uploader $u$ in class $c_{i}$, $K_u$ is the number of classes touched by $u$, and $\alpha$ and $\beta$ are tunable weights to set the relevance of each term, both $\in [0,1]$.
The first term is the amount of data in $c_{i}$ by $u$, whereas the second term is the average number of labels per class, accounting for the scattering of $u$ across classes.
Uploaders are sorted in ascending score order and the content of low-score uploaders is transferred first.
By tuning $\alpha$ and $\beta$ we aim to promote the uploaders providing a small amount of data in the class under question, $c_{i}$, with minimal or no scattering. 
This facilitates the adjustment to a target class distribution while minimizing contamination (both WC and BC).
This first group of uploaders is followed by others with smooth scattering across classes, avoiding uploaders with large contributions concentrated in specific classes.
This again facilitates adjusting to a target distribution while minimizing the need to split content from the same uploader in one class (i.e., WC contamination), but allowing BC contamination.

Once the validation set is initialized and the uploaders are sorted per-class, we allocate data to the validation set as shown in Algorithm \ref{alg:val_split}.
%A pseudocode version of the process is shown in Algorithm \ref{alg:val_split}.
\vspace{-2mm}
\begin{algorithm}[h]
\label{alg:val_split}
\DontPrintSemicolon
\SetAlgoLined
\KwData{Initialized validation data per-class $V=\left\lbrace v_{c_{i}} \right\rbrace_{i=1}^C$, uploaders ranking in development set per-class $U=\left\lbrace \boldsymbol{u_{c_{i}}} \right\rbrace_{i=1}^C$}
 %initialization\;
 \For{pass $n = 1,2, \dotsc  N $}{
    \For{class $c_i \in C$}{
      get current validation target $t_{c_{i}}$ \;
      \While{$v_{c_{i}} < t_{c_{i}}$}{
        get next uploader $u$ in ranking $\boldsymbol{u_{c_{i}}}$ \;
        $v_{c_{i}} \leftarrow v_{c_{i}}$ + data from $u$ in $c_{i}$ \;
        \If{data is multilabel to class $c_j$}{
           $v_{c_{j}} \leftarrow v_{c_{j}}$ + data from $u$ in $c_{j}$ \;
        }
      }
    }
 }
 \KwResult{A candidate validation set}
 \caption{Data allocation to validation set}
\end{algorithm}
\vspace{-1mm}
We traverse the classes in several passes, and, for each class $c_i$, we progressively allocate content from the ranked uploaders until a target data amount $t_{c_{i}}$ is reached.
Note that when separating the class $c_i$, the algorithm does not care about a given uploader $u$ contributing to another class $c_j$ (BC contamination), unless there is at least one clip bearing labels for both $c_i$ and $c_j$.
WC contamination can be produced in lines 6 and 8.
We designed the step in line 6 so that, if adding the content from $u$ implies exceeding the validation target $t_{c_{i}}$ by more than 15\%, two things can happen.
If the current validation amount is $v_{c_{i}} > 0.75 t_{c_{i}}$, the content is not transferred, $v_{c_{i}}$ is deemed sufficient and the procedure halted for $c_i$. 
This flexibility allows the minimization of WC contamination at the expense of deteriorating stratification.
Else, if $v_{c_{i}} <= 0.75 t_{c_{i}}$, the content from $u$ in $c_i$ is split and the amount needed to reach $t_{c_{i}}$ is allocated, causing WC contamination. 
Similar heuristics are adopted for line 8.

Using the proposed scheme, we process clips from the 4936 uploaders in the development set.
Due to the high variability of users, the process needs initial debugging with a subset of classes in order to tune the weights $\alpha$ and $\beta$. 
We finally use $\alpha=0.4$ and $\beta=0$ when an uploader contributes only to one class, and $\alpha=0.3$ and $\beta=0.7$ otherwise.
We use $N=2$ passes starting from classes in need of more validation data, which allows us to reach a reasonable stratification. 
The target validation proportion is 15\% of the development labels per-class, except for the largest 17 classes where we reduced this percentage progressively.
%(as we considered that enough data was allocated).
The first-pass target is to fill 60\% of the 15\%-target, which is the goal in the second-pass.
We only consider the leaf nodes for this process ($C=144$).
This is done for simplicity and because the leaf nodes are the most specific data that will receive labels from the rest of the ontology levels upon propagation to their ancestors.
In this way, validation data at all levels of the ontology is guaranteed.

The outcome is a validation set which represents a tradeoff between stratification and contamination.
%It is composed of 4170 audio clips (of which 158 do not have leaf labels, and ended up due to the step of line 8 in Algorithm 1 (multilabel nature to non-leaf nodes).
Composed of 4170 audio clips, it amounts to 13.3\% of the content associated with leaf nodes and 10.2\% of the entire development set.
Its main statistics are listed in Table~\ref{tab:val_set}.
Out of the 2224 uploaders with content in the validation set, 641 also have content in the train set---mostly corresponding to BC contamination.
Sec. \ref{ssec:exp_val} describes sound event tagging experiments comparing the proposed split to others obtained via off-the-shelf approaches.
% This candidate split is the result of a number of design choices.
% However, other choices might be desirable (e.g., proportion, contamination, usage of intermediate nodes, etc.) depending on researchers' needs.
% Alternative validation sets can be created using the clip metadata provided in FSD50K, which includes uploader information.
\begin{table}[ht]
\vspace{-6mm}
\caption{Main statistics for candidate validation set}
\vspace{-2mm}
\centering
\begin{tabular}{cccc}
\toprule
\textbf{clips} & \textbf{duration} & \textbf{uploaders} \\
\midrule
4170             & 9.9h           & 2224           \\
%\midrule
\bottomrule
\end{tabular}
\label{tab:val_set}
\vspace{-3mm}
\end{table}

\vspace{-3mm}
\subsection{Hierarchical Label Propagation}
%=========================================================================
At this point, the labels in train, validation and evaluation sets are usually from classes corresponding to lower levels of the ontology, especially for the evaluation set (see Sec. \ref{ssec:task_gen}).
To obtain an exhaustive labelling hierarchy-wise, we need to propagate the current labels in the upwards direction to the root of the ontology, determining the ancestors in the hierarchical path and automatically assigning them to the corresponding audio clips.
This hierarchical label propagation process is sometimes referred to as \textit{label smearing} \cite{gao2017knowledge,hershey2021benefit}.
%\TODO{this could be provided with the dataset, or in Zenodo}
In most cases, this is straightforward as there is one single unequivocal path from a given low-level node to the root.
However, in other cases, nodes and root are connected by more than one path.
Among these multiple-path cases, some have all the paths valid by default according to the semantics of the node.
This allows straightforward propagation as in the single-path case, e.g., \textit{Doorbell} can be directly propagated to \textit{Door} and \textit{Alarm}.
However, in the majority of cases, only a subset of the paths is valid (often only one path), or even none of the paths is valid by default due to the parents-node relationship. For instance, \textit{Buzz} cannot be directly propagated to its parents \textit{Fly, housefly} or \textit{Bee, wasp, etc.} unless we have explicit information about the source of the buzz sound. 
In these cases, we need knowledge of the correct immediate parent(s) to unambiguously infer ancestors for a complete hierarchical labelling.
Parents disambiguation can be carried out in different ways depending on the annotation task.
In the clips annotated only with the validation task, the disambiguating parents will exist if and only if the nomination system proposed them.
For the clips annotated also with the refinement task, raters were instructed to specify the disambiguating parents when needed; however, 
%this additional annotation subtask is an added complexity that rarely happens, which makes it prone to errors. Upon inspection of the gathered annotations, 
we detected that they were not always specified.

As a result, in these cases, ancestors cannot be inferred from the leaf node, leading to hierarchical paths featuring missing parts.
For example, \textit{Growling} is connected directly to \textit{Animal} in several cases where information of the source animal is not available.
The policy followed in case of ambiguous ancestors was to not include these labels (hence potentially creating missing “Present” labels in the mid- or high-levels of the ontology) instead of possibly generating incorrect labels.
In the development set, these cases are provided \textit{as is} since it is less critical. 
By contrast, because the cases in the evaluation set are more critical, they were partially reviewed and corrected. 
% The potential impact of missing intermediate nodes is restricted to some instances of class labels with multiple-paths where the disambiguating parents could not be determined, and thus we expect this to have a minimal impact.

To finalize this process, we filter out labels beyond the 200 selected. In the majority of cases, these correspond to abstract or blacklisted classes of the ontology \cite{gemmeke2017audio}. This is another reason why some clips have labels up to the ontology root while others only have a portion of the ancestors or even one single label. For example, \textit{Whoosh, swoosh, swish} has no hierarchy as all class labels in its path were either removed previously due to specified constraints (\textit{Arrow}) or removed in this last step (as classes above \textit{Arrow} are abstract). This can be easily spotted in the provided ground truth CSV files.\textsuperscript{\ref{foot_zenodo}}
%==
The number of labels before/after the hierarchical label propagation process can be seen in Table~\ref{tab:main_stats} (\textit{unpropagated} and \textit{propagated}, respectively).
The outcome is a set of hierarchically propagated labels consistently encompassing all relevant levels of the ontology.
%, through upwards propagation of the initially human-provided lower-level class labels. 
Note the considerable increase of labels, despite that we are ignoring parts of the ontology. 
This is the final ground truth provided for FSD50K.